\documentclass{aa}  

\usepackage{txfonts}
\usepackage{natbib}
\usepackage{graphics}
\usepackage{graphicx}
\usepackage{xcolor}
\usepackage{amssymb}
\usepackage{amsmath}
\usepackage{multicol}
\usepackage{subfig}
\bibpunct{(}{)}{;}{a}{}{,}
\begin{document}

   \title{Aggregates of clusters in the Gaia data}

   \author{M.~Piecka \and E.~Paunzen}
   
    \institute{Department of Theoretical Physics and Astrophysics, Masaryk University,
    Kotl\'a\v{r}sk\'a 2, CZ-611\,37 Brno, Czech Republic
    \email{408988@mail.muni.cz}
    }

   \date{Received X XX, XXXX; accepted X XX, XXXX}

 
  \abstract
   {The precision of the parallax measurements by Gaia is unprecedented. As of Gaia Data Release 2, the number of known nearby open clusters has increased. Some of the clusters appear to be relatively close to each other and form aggregates, which makes them interesting objects to study.}
   {We study the aggregates of clusters which share several of the assigned member stars in relatively narrow volumes of the phase space.}
   {Using the most recent list of open clusters, we compare the cited central parallaxes with the histograms of parallax distributions of cluster aggregates. The aggregates were chosen based on the member stars which are shared by multiple clusters.}
   {Many of the clusters in the aggregates have been assigned parallaxes which coincide with the histograms. However, clusters that share a large number of members in a small volume of the phase space display parallax distributions which do not coincide with the values from the literature. This is the result of ignoring a possibility of assigning multiple probabilities to a single star. We propose that this small number of clusters should be analysed anew.}
   {}


   \titlerunning{}
   \authorrunning{}
   \maketitle
%
\section{Introduction}\label{section:1}

Assuming that the stellar members of a given open cluster are born from a molecular cloud which is chemically almost homogeneous, one can determine the distance, reddening, age, and metallicity for these stars with the help of statistical methods (for example, isochrone fitting techniques). The latter two quantities are especially difficult to determine for single stars. This shows that galactic star clusters represent an important tool for studying the evolution and properties of stars and the Milky Way itself.

Binary clusters are especially interesting for studying the formation and evolution on a
global scale. The pairs could either be formed at the same time \citep{Priyatikanto2016} or sequentially caused by stellar winds or supernova shocks generated by one cluster inducing the collapse of a nearby cloud, thus triggering the formation of a companion cluster \citep{Goodwin1997}. Furthermore, they can be formed completely separately and then captured either by tidal forces \citep{vdBergh1996} or resonant trapping
\citep{Dehnen1998}. 

Since its first data release, the Gaia satellite presented an improvement in the measured parallaxes of stars and expanded the list of known stars by several orders of magnitude. With the use of Gaia's most recent accurate astrometric and photometric measurements \citep{GaiaDR2}, it is not only possible to study clusters at larger distances but also to refine the list of nearby clusters. \citet[][CG18 from now on]{Cantat2018} present a status report for 1229 open cluster based on the Gaia Data Release 2 (Gaia DR2). They establish the parallaxes, the proper motions, and the most likely distances of the clusters, together with the membership probabilities of the individual stars (based on UPMASK). These data were later expanded and currently describe 1481 clusters \citep[][CG20]{Cantat2020}.

\begin{table*}[ht]
\caption{Aggregates used for the detailed analysis. The presented values are the mean values for the whole clusters taken from CG20.}
\label{table-1}
\centering
\begin{tabular}{c | l r r r r r}
\hline\hline
Aggregate & Cluster & $l$ [deg] & $b$ [deg] & $\varpi$ [mas] & $\mu_{\alpha}$ [mas/yr] & $\mu_{\delta}$ [mas/yr] \\
\hline
    Agg02 & ASCC 19 & 204.914 & -19.438 & 2.768 & 1.152 & -1.234 \\
         & Gulliver 6 & 205.246 & -18.138 & 2.367 & -0.007 & -0.207 \\
         & UBC 17a & 205.335 & -18.019 & 2.753 & 1.582 & -1.200 \\
         & UBC 17b & 205.142 & -18.179 & 2.376 & 0.078 & -0.163 \\
    \hline
    Agg07 & Alessi 5 & 288.058 & -1.966 & 2.501 & -15.411 & 2.503 \\
         & BH 99 & 286.585 & -0.592 & 2.225 & -14.494 & 0.919 \\
    \hline
    Agg17 & COIN-Gaia 40 & 174.048 & -0.794 & 0.470 & 0.393 & -2.762 \\
         & Gulliver 53 & 173.226 & -1.141 & 0.384 & 0.401 & -2.837 \\
         & Kronberger 1 & 173.106 & 0.049 & 0.443 & -0.050 & -2.199 \\
         & NGC 1893 & 173.577 & -1.634 & 0.267 & -0.231 & -1.410 \\
         & Stock 8 & 173.316 & -0.223 & 0.446 & 0.094 & -2.249 \\
    \hline
    Agg32 & FSR 0686 & 156.835 & -2.212 & 1.087 & -1.088 & -2.556 \\
         & UBC 55 & 156.823	& -2.190 & 1.079 & -1.067 & -2.604 \\
    \hline
    Agg42 & Gulliver 56 & 185.664 & 5.915 & 0.458 & 0.534 & -3.240 \\
         & UBC 73 & 185.661 & 5.918 & 0.448 & 0.455 & -3.210 \\
    \hline
    Agg53 & RSG 7 & 108.781 & -0.320 & 2.337 & 4.927 & -1.865 \\
         & RSG 8 & 109.150 & -0.484 & 2.210 & 5.343 & -1.654 \\
\hline
\end{tabular}
\end{table*}

\citet{Soubiran2018} examined the kinematical properties in the 6D phase space of open clusters derived from Gaia DR2 data. They have confirmed that the velocity distribution of clusters coincides well with the velocity distribution of stars at given Galactic locations. Furthermore, they have shown that several clusters seem to be confined to the same (small) volume of the phase space. This suggests that there may be a physical connection between the clusters.

Any procedure used for detecting clusters and deriving cluster membership probabilities is expected to have some limits. Using only astrometric measurements (stellar coordinates, proper motion, and parallax) necessarily leads to the assignment of non-member stars to the clusters, although Bayesian statistics may help to constrain the number of such false identifications. For the parallax space, this is discussed in \citet{Luri2018}, for example. Additional information can be extracted from the colour-magnitude diagrams of the clusters, which may further constrain the membership probabilities. Overall, the final list of cluster members will never completely resemble the true host cluster -- the quality of such a representation will depend on the number of detected cluster members (signal), the number density of the surrounding field stars (noise), and instrumental errors (random and systematic).

Based on the arguments above, it is possible for an automated procedure to fail under special circumstances. An example would be a case of two apparently (or physically) neighbouring clusters which may overlap in the position, parallax, and proper motions space -- this was already predicted by the creators of the UPMASK method \citep{KroneMartins14}. The only assumption used by this procedure is that the distribution of stars in the phase space is more tight than in the case of a random (uniform) distribution. The procedure is composed of two main steps -- the identification of clusters in the data (based on the \textit{k}-means clustering method) and the comparison of the groupings with random distributions (returning a binary value which states whether the grouping is a cluster or not). The process is applied several times, and the membership probability of each star is determined based on the ratio $\frac{\textrm{number of positive assignments}}{\textrm{total number of iterations}}$.

If some of the cluster members are assigned to more than one cluster, one must be cautious with the calculation of the cluster parameters. We find it important to re-examine the data presented by CG20. Our goal is to investigate the number of duplicates appearing in the data set and determine their impact on the derived cluster parameters. However, we do not aim to search for new cluster groups.

\begin{figure*}
   \centering
	 \begin{tabular}{ll}
   \multicolumn{2}{c}{\subfloat{\includegraphics[scale=0.5]{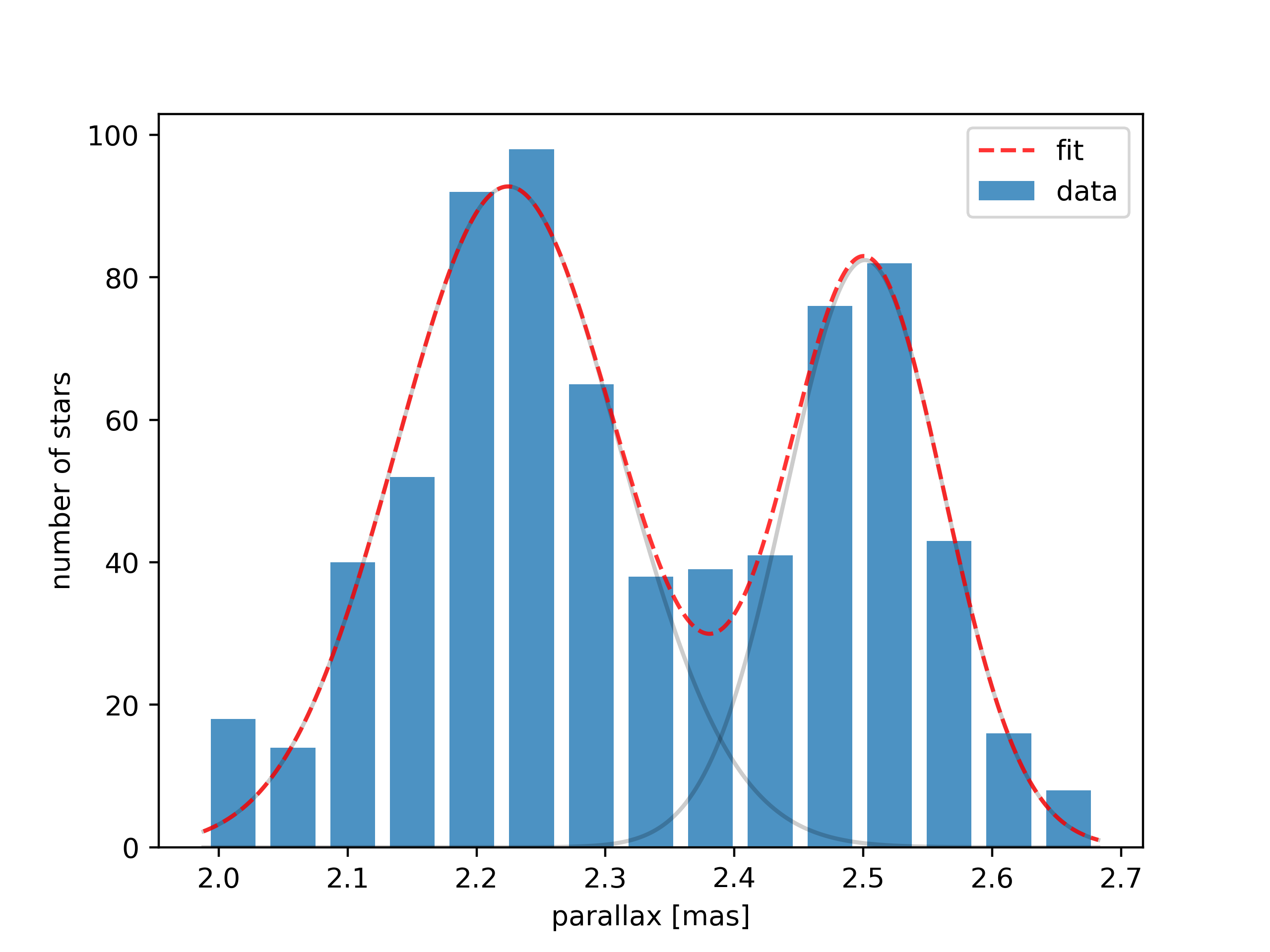}}} \\ 
   \subfloat{\includegraphics[scale=0.5]{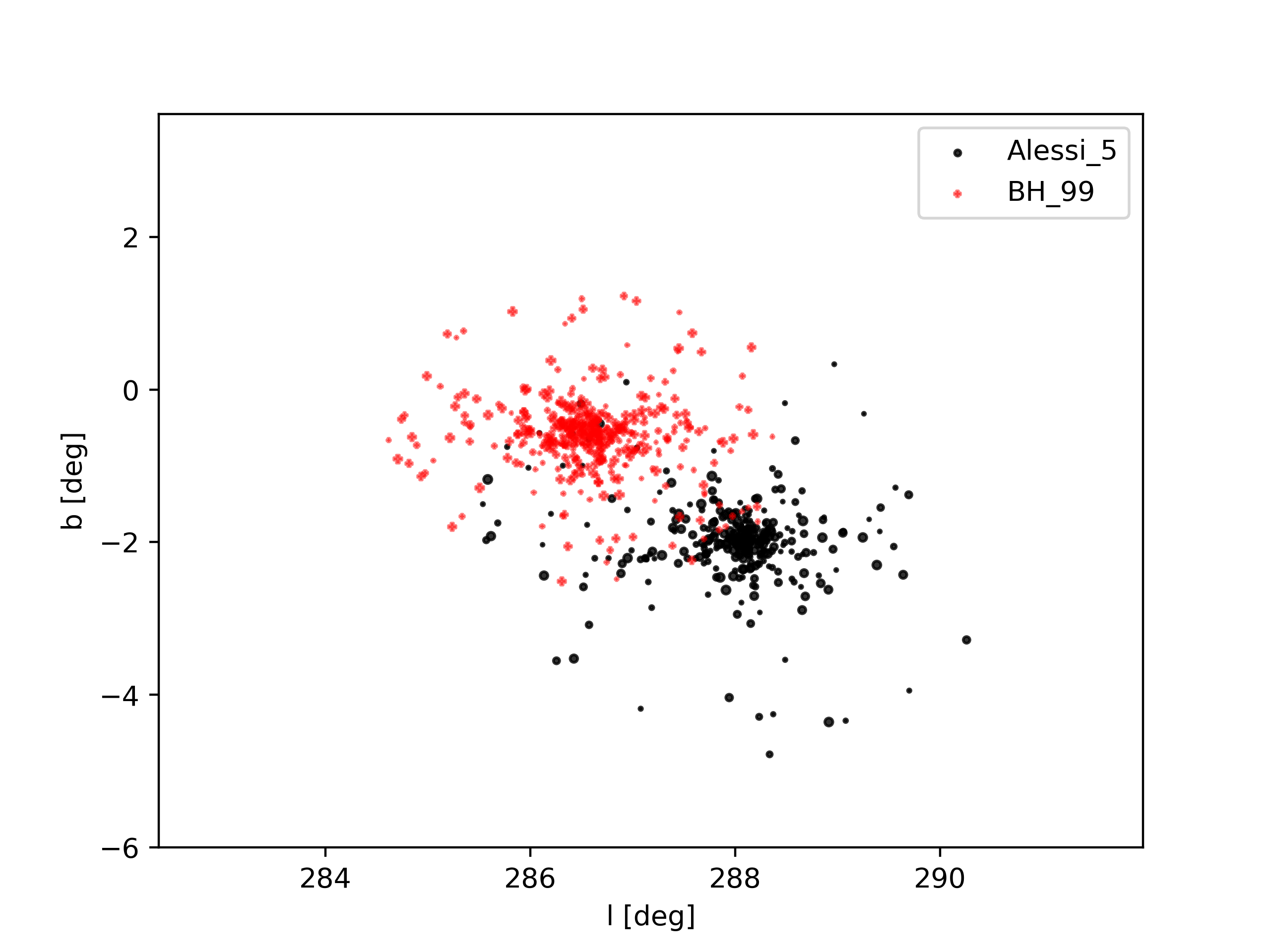}} \quad &
	 \subfloat{\includegraphics[scale=0.5]{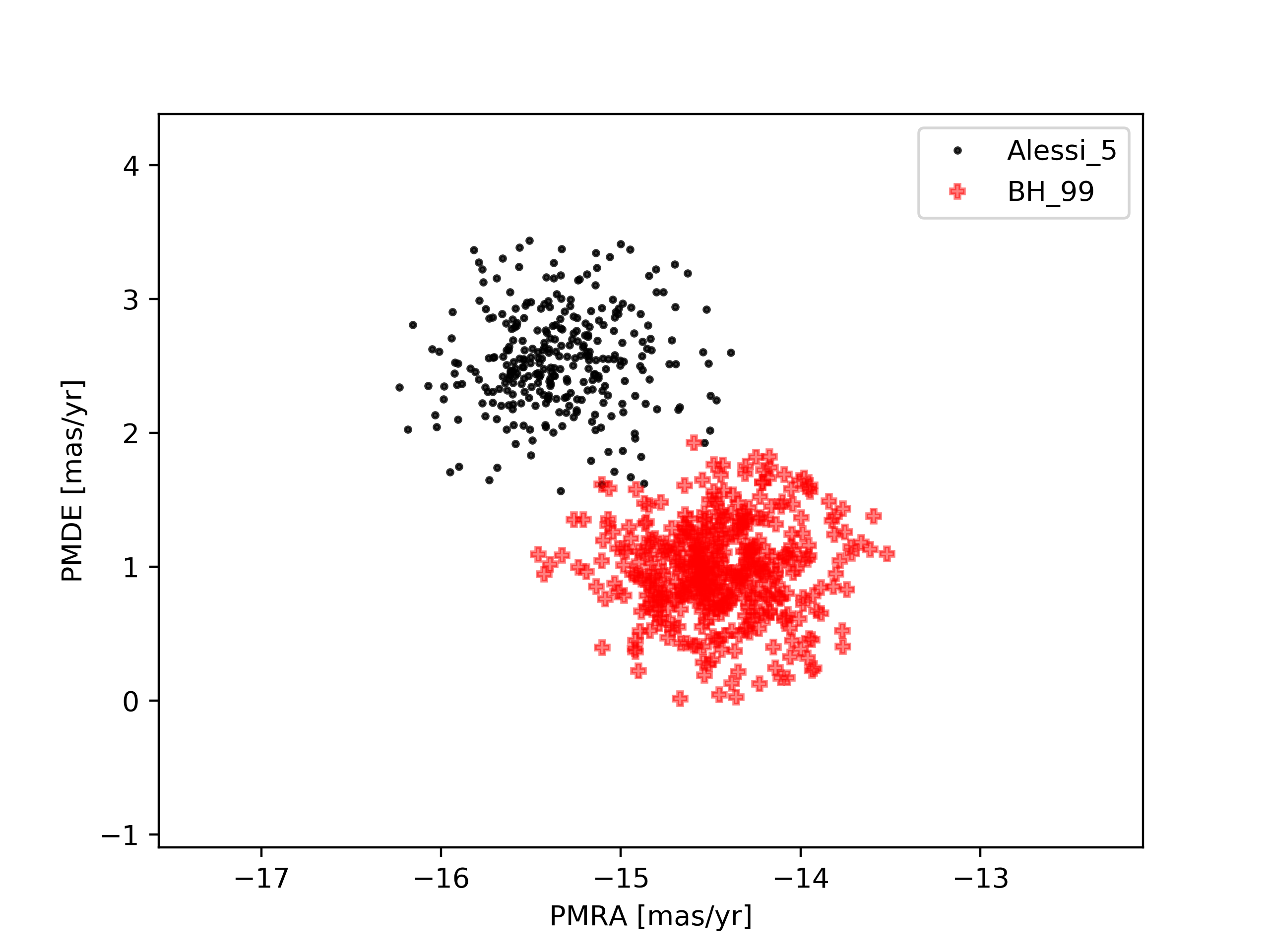}} \\
   \subfloat{\includegraphics[scale=0.5]{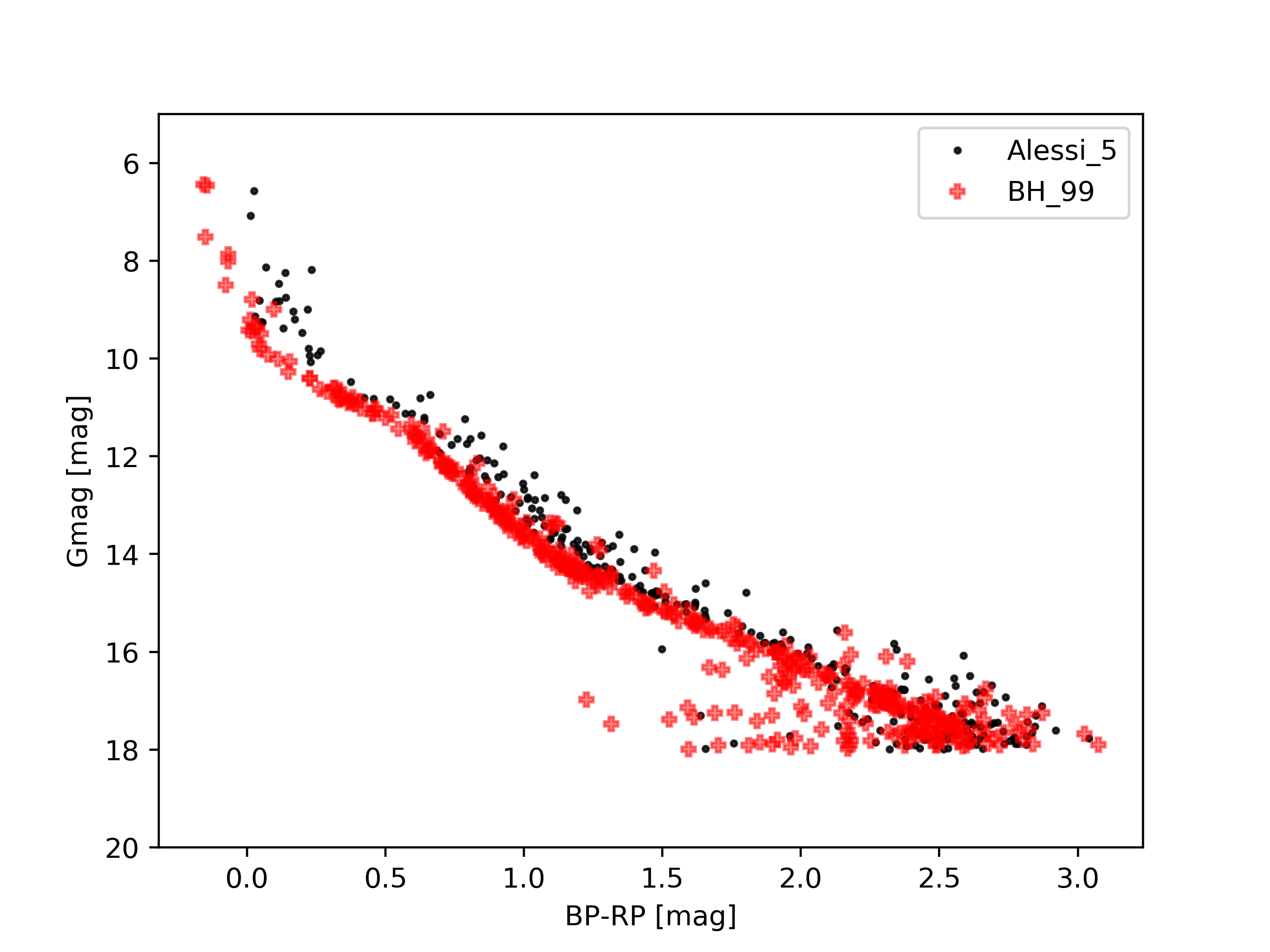}} \quad & 
   \subfloat{\includegraphics[scale=0.5]{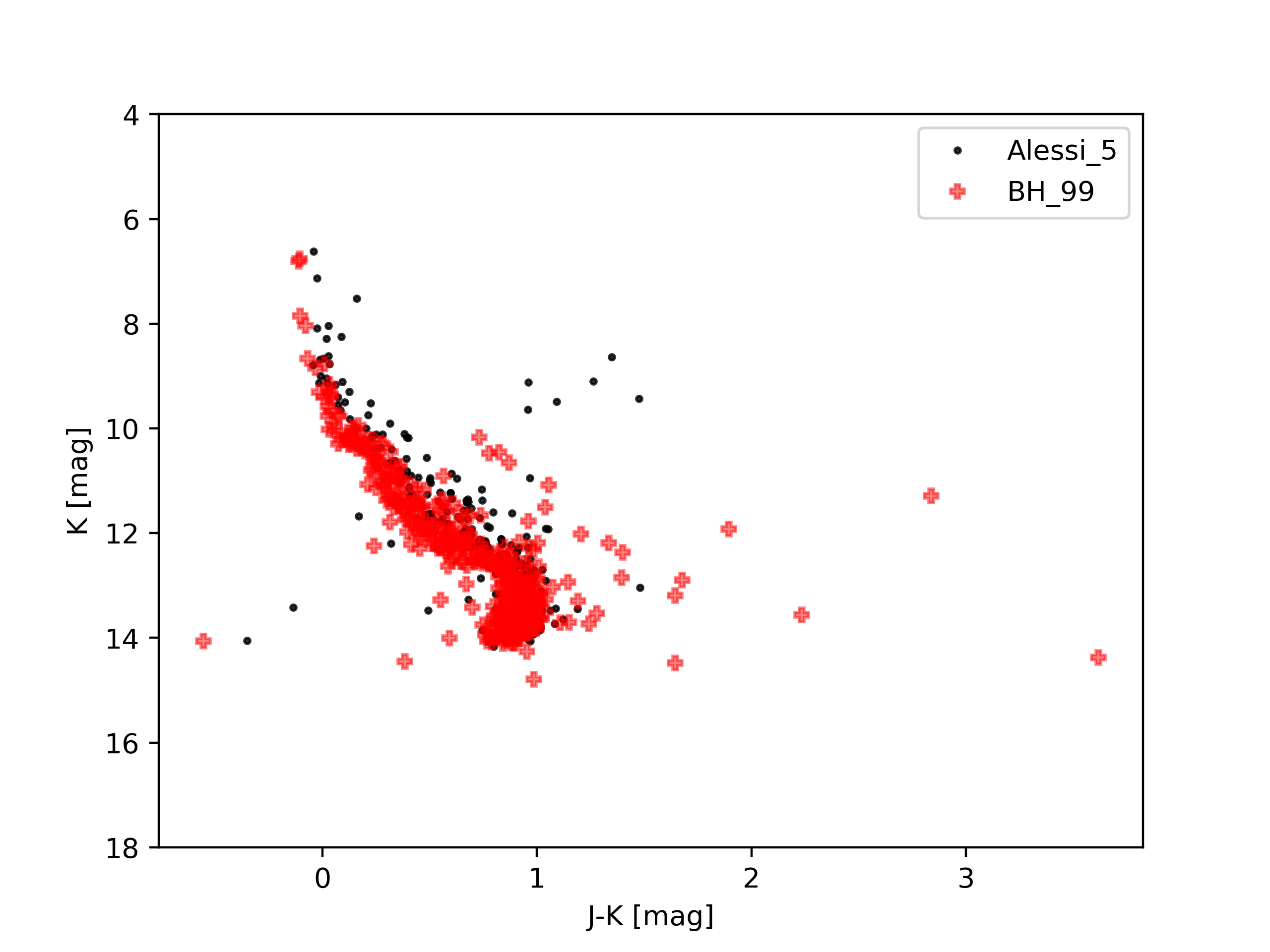}} \\
	 \end{tabular}
   \caption{Different slices through the phase space of Agg07 (Alessi~5, BH~99), together with the complete colour-magnitude diagrams (based on Gaia and 2MASS photometry). Top: the histogram of parallaxes, excluding the duplicate cluster members. The best fit was achieved with the following double-Gaussian function parameters: $\varpi_1 = 2.225 \pm 0.007$ mas, $\sigma_1 = 0.087 \pm 0.007$ mas, $\varpi_2 = 2.501 \pm 0.006$ mas, $\sigma_2 = 0.061 \pm 0.006$ mas. Middle-left: coordinates of the stars in Agg07. Size of the points indicates values of the observed magnitude $G$. Middle-right: proper motion diagram of Agg07. Bottom-left: colour-magnitude diagram based on the observed values of $B_P-R_P$ and $G$). Bottom-right: colour-magnitude diagram based on 2MASS photometry. The stars were located using the coordinates in Gaia database. The two clusters remain distinguishable in this photometric system.}
   \label{phasespace_1}
\end{figure*}

\section{Data selection}\label{section:2}

The data set from CG20 consists of two data tables. The list of clusters (1481 entries) includes the number of members for each cluster, derived positions, proper motions, and parallaxes. The list of cluster members (435\,833 entries), for which the measured Gaia values of astrometric and some photometric measurements are available, presents the derived membership probabilities and IDs of the host clusters.

Based on the IDs of member stars, we have found 133 clusters containing stars which are also assigned to at least another cluster. The occurrence of these 'duplicates' (2148 individual stars) is the sign of an overlap in the reduced phase space (leaving out radial velocities) in which the determination of the membership probabilities was executed. This overlap is mostly attributed to the measurement uncertainties. Very rarely, this could also be a hint that the host clusters may be physically close to each other and/or even overlapping. In such cases, the determination of the membership probability requires a more complex approach.

For the probability of a given star being a member of at least one of the assigned clusters, the sum of the probabilities for all assigned clusters must be considered. We find that 1011 of the 2148 duplicate stars have summed probabilities of $p_{\textrm{sum}}>1.0$ (of these, 456 stars with $p_{\textrm{sum}}>1.5$ and one star with $p_{\textrm{sum}}>2.0$). Clearly, the approach used for the membership probability evaluation does not take a possible assignment to several clusters into account. For the purpose of our analysis of some of these cluster aggregates, we simply assume that all stars assigned to the aggregates (without duplicity) are members.

We estimate 60 aggregates from the sample of 133 clusters (Table~\ref{table-A}). This was done with the use of transitivity -- if cluster $A$ shares a member with cluster $B$, and if cluster $B$ shares a member with cluster $C$, then all three clusters form an aggregate. The found aggregates can be distinguished based on the number of included clusters. In 52 instances, the aggregate consists of only two clusters. There are also four aggregates composed of three clusters, three aggregates containing four clusters, and one case of a five-cluster aggregate.

We have chosen an example of double-Gaussian distributions in the parallax space (Agg07), which serves as a reference point. There are at least four aggregates (Agg02, Agg32, Agg42, and Agg53) for which the constituent clusters overlap in the whole phase space in such a way that they are very hard to distinguish from each other. These were chosen in such a way that at least two clusters follow these rules: The difference between the median values of their proper motions (each component) is less than half the sum of the standard deviations; the same was applied for the positions; and for parallaxes, the full sum of the standard deviations was used. Moreover, the five-component aggregate (Agg17) was also considered. The chosen clusters are summarised in Table~\ref{table-1}.

Some of the stars presented in the data were assigned to the same cluster twice -- such cases are highlighted by our procedure but not considered to be actual duplicates in our analysis. This situation occurred in the case of UBC~3, UBC~19, and UBC~31. For example, \texttt{Gaia DR2 4505874688732436992} is presented twice for the host cluster in the \texttt{VizieR} catalogue \texttt{J/A+A/633/A99/members} (CG20).

\section{Case analysis}\label{section:3}

In this section, we present an analysis of the chosen aggregates. We are especially interested in comparing the mean parallaxes derived in the data source with the parallax histograms of the whole aggregates. We predict that the derivation of parallaxes should be severely affected for aggregates in which multiple stars are being shared by the clusters.

The histograms (points located at the centre and maximum height of each bin) were fitted by a sum of two Gaussian functions -- this type of model has six free parameters (peak height, central position, and width of both peaks). This was achieved with the use of the basic \texttt{scipy.optimize.curve\_fit} function available in \texttt{Python}.

For the comparison of cluster parameters (age, extinction, and metallicity) of the individual objects, we used the results from \citet{Dias2002}, \citet{Kharchenko2013}, and \citet{Bossini2019}. Most of the figures related to the analysis are presented in Appendix \ref{section:Appendix}.

\subsection{Alessi~5 and BH~99}

Alessi~5 was first mentioned in the work by \citet{Dias2002}. It is a cluster of intermediate age ($\log t = 7.723$) with a good amount of extinction being present in the line of sight ($A_V = 0.592$ mag). The cluster appears to be somewhat metal-poor, with $\textrm{[Fe/H]}=-0.133$.

BH~99 is almost as old as Alessi~5 ($\log t = 7.908$). However, the estimated extinction appears to be somewhat lower ($A_V = 0.203$ mag). The metallicity of this cluster has been assumed to be solar, $\textrm{[Fe/H]}=0.000$.

The comparison of the parallaxes from Table~\ref{table-1} with the histogram (fitted by a double-Gaussian) of this aggregate (Fig.~\ref{phasespace_1}, top panel) clearly shows that the parallaxes were derived correctly. The clusters share only one of the total number of 722 stars.

Based on the parameters listed in Table~\ref{table-1}, it seems that these clusters are fairly close to each other. Using the Galactic coordinates listed in \texttt{VizieR}, their relative distance is found to be about 50 pc. However, when compared with the clusters analysed below, these two are fairly distant in the position on sky and noticeably differ in their mean proper motion (Fig.~\ref{phasespace_1}, middle panels).

This aggregate is mentioned in \citet{Soubiran2019}. Due to the large difference in metallicities, the most likely assumption is that these clusters represent two distinct groups of stars, despite their proximity in the phase space. The difference can be clearly seen in the colour-magnitude diagrams of the aggregate (Fig.~\ref{phasespace_1}, bottom panels, adopting the membership probabilities from the source data). However, differential extinction may also play a role in the determination of the cluster parameters -- given their relative distance, the difference between the derived extinction values seems to be quite large ($\Delta A_V \sim 0.25$ mag over $\sim 50$ pc).

\subsection{FSR~0686 and UBC~55}

The age of FSR~0686 is not specified in \citet{Bossini2019}. \citet{Dias2002} provided a value of $\log t = 8.610$, making it a fairly old cluster. Curiously, it is discussed in CG18 that the angular diameters of FSR clusters differ when looking at the catalogues from Dias and Kharchenko \citep{Kharchenko2013}.

UBC~55 is one of the newly discovered clusters, which was first presented in the source data used in this work. For this reason, its cluster parameters have not been determined yet.

The histogram of parallaxes is displayed in the top panel of Fig.~\ref{phasespace_2} -- the source parallaxes slightly differ from those determined here. However, given the relatively small stellar count, fitting a single Gaussian function is also possible, leading to a parallax value which matches those from the literature.

The distance between these clusters was calculated to be less than 10 pc. Their position and proper motion appear to be almost identical (Fig.~\ref{phasespace_2}, middle panels). The situation is the same when we look at the colour-magnitude diagrams (Fig.~\ref{phasespace_2}, bottom panels). In principle, this makes the aggregate a good candidate for studies of binary clusters. However, 42 out of 80 stars (all have $p_{\textrm{sum}}>1.0$) are being shared between the clusters, which makes it difficult to prove whether this aggregate truly consists of two different groups.

\subsection{Gulliver~6 and UBC~17b}

The stars in Agg02 seem to form two different subgroups, which can be seen in all planes of the phase space and even in the colour-magnitude diagrams (Fig.~\ref{phasespace_3}). ASCC~19 and UBC~17a seem to be fairly similar when looking at the different plots of the clusters. However, they remain relatively distinguishable in the coordinate and proper motions space. This subgroup is significantly far away in the sky from the second subgroup and has a different proper motion (similar to the case of Alessi~5 and BH~99).

Gulliver~6 and UBC~17b share a very small volume of the phase space. For this reason, we have focused our attention on this subgroup. Literature provides no age or metallicity estimates for these two clusters. They share 75 out of 412 stars, with 103 being assigned to UBC~17b. Similar to the previous aggregate, most of the members of one cluster were also derived as members of the other cluster, all of them having a summed probability larger than $1.0$. We do not see a clear distinction between the clusters in the parallax distribution (Fig.~\ref{phasespace_3}, top panels) -- the two clusters can be easily fitted with a single Gaussian function.

UBC~17a and UBC~17b are very close to each other in the coordinates space. However, they become quite distinguished when looking at their parallaxes and proper motions.

Gulliver~6 has been previously compared to ASCC~19, and several other clusters, in the phase space. \citet{Soubiran2018} determined that the aggregate consists of five clusters. We must note that our requirements for the phase space were more strict -- we only searched for the duplicates in the data. This led to the exclusion of ASCC~16, ASCC~21, and NGC~2232 from our list.

\begin{figure*}
   \centering
	 \begin{tabular}{ll}
   \subfloat{\includegraphics[scale=0.5]{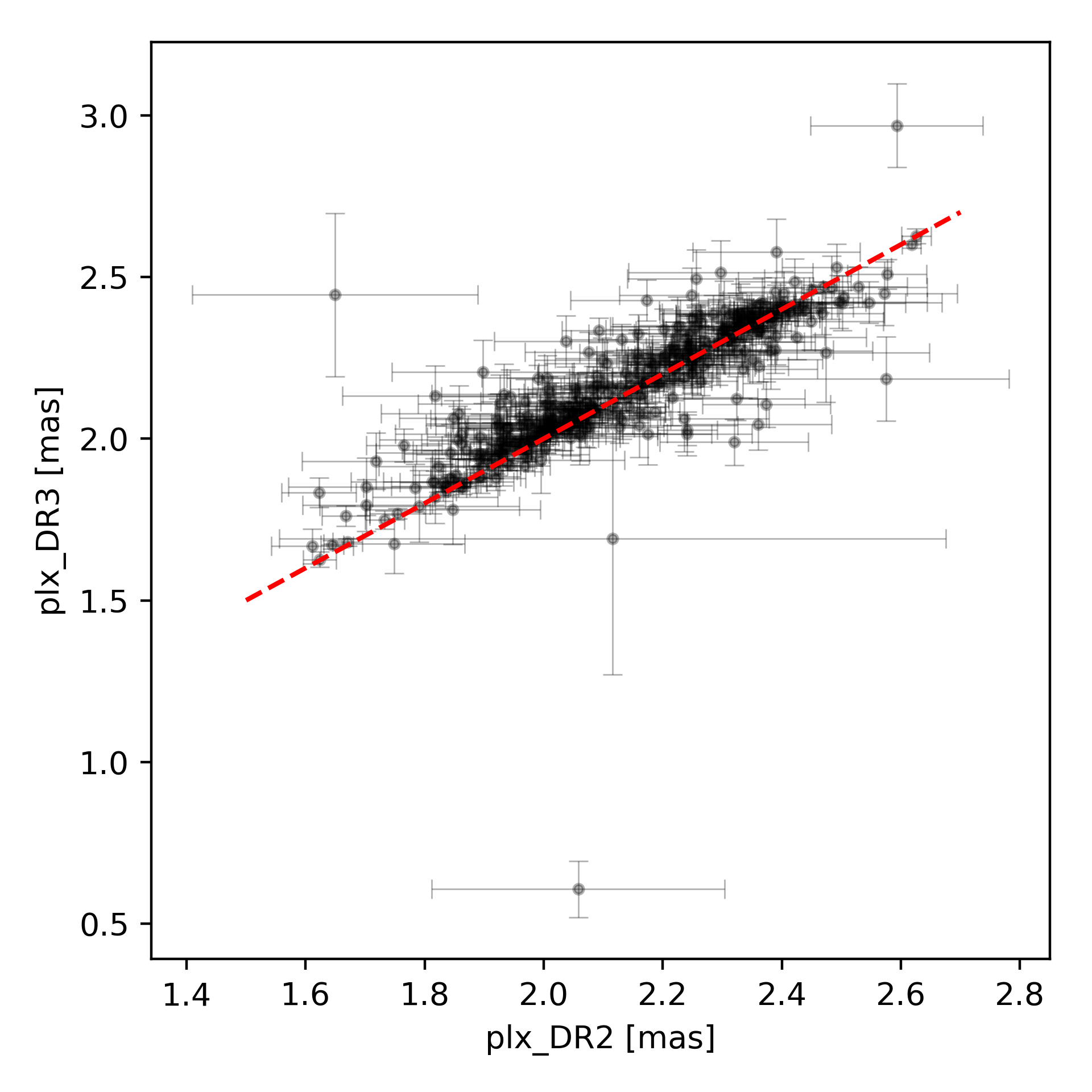}} \quad &
	 \subfloat{\includegraphics[scale=0.5]{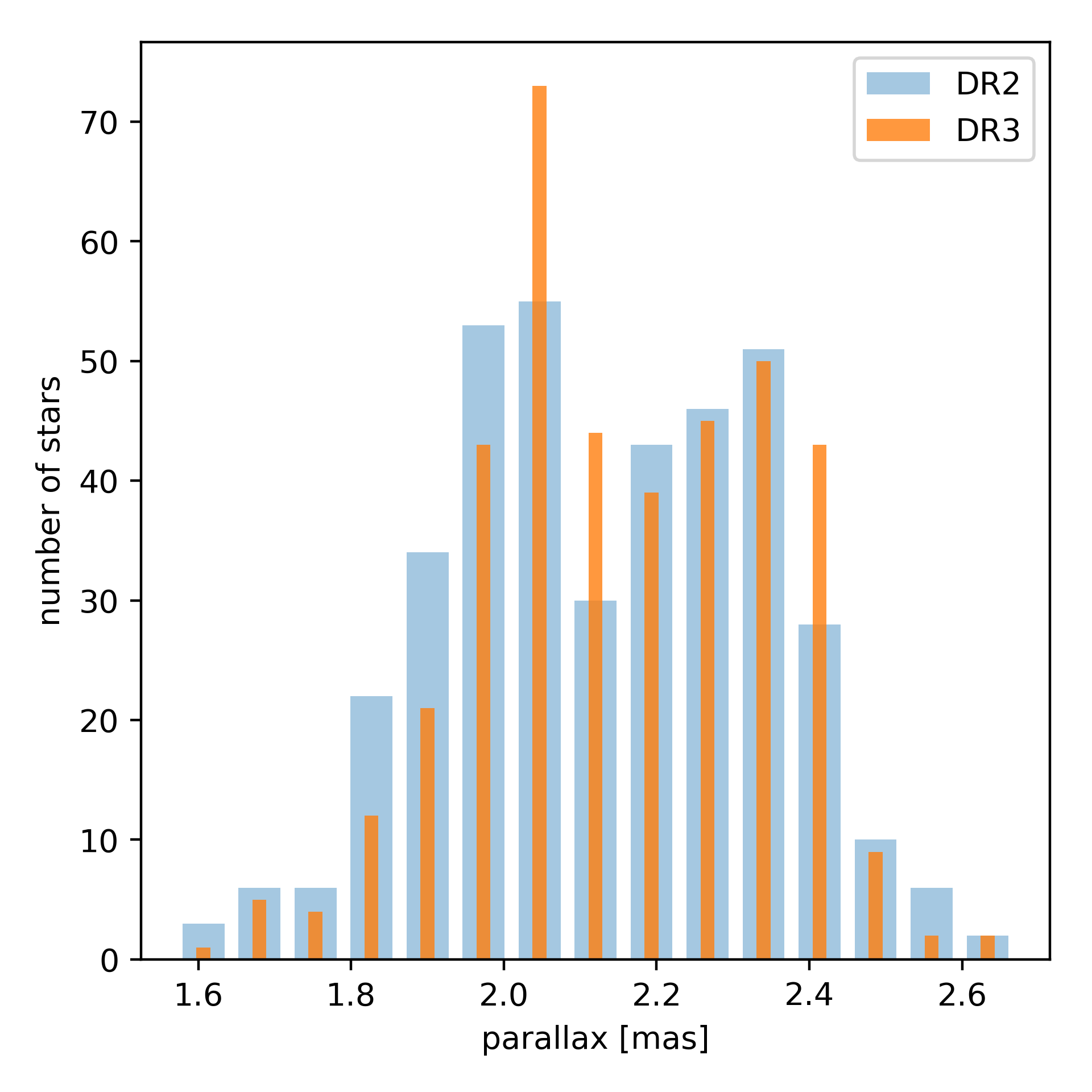}} \\
   \subfloat{\includegraphics[scale=0.5]{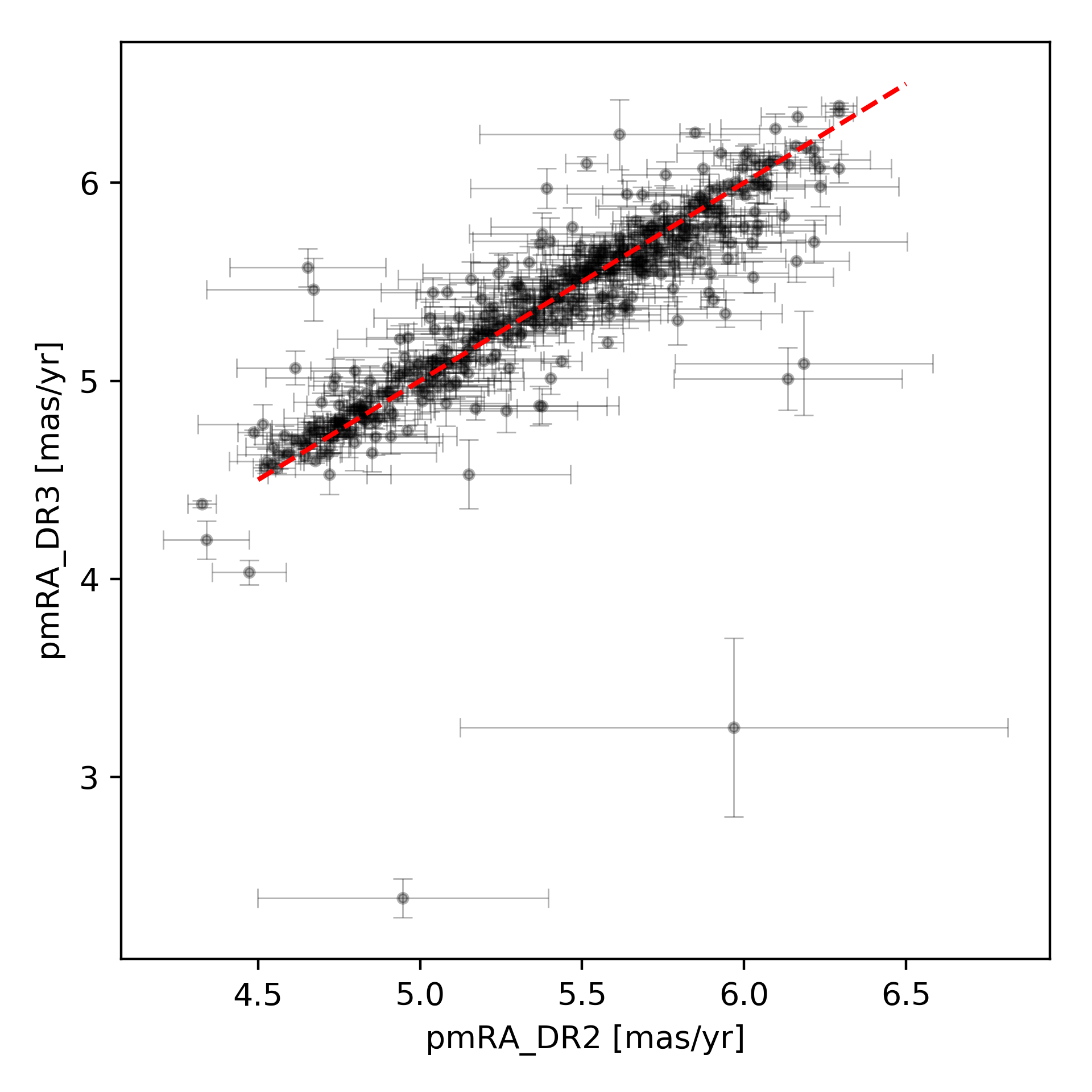}} \quad & 
   \subfloat{\includegraphics[scale=0.5]{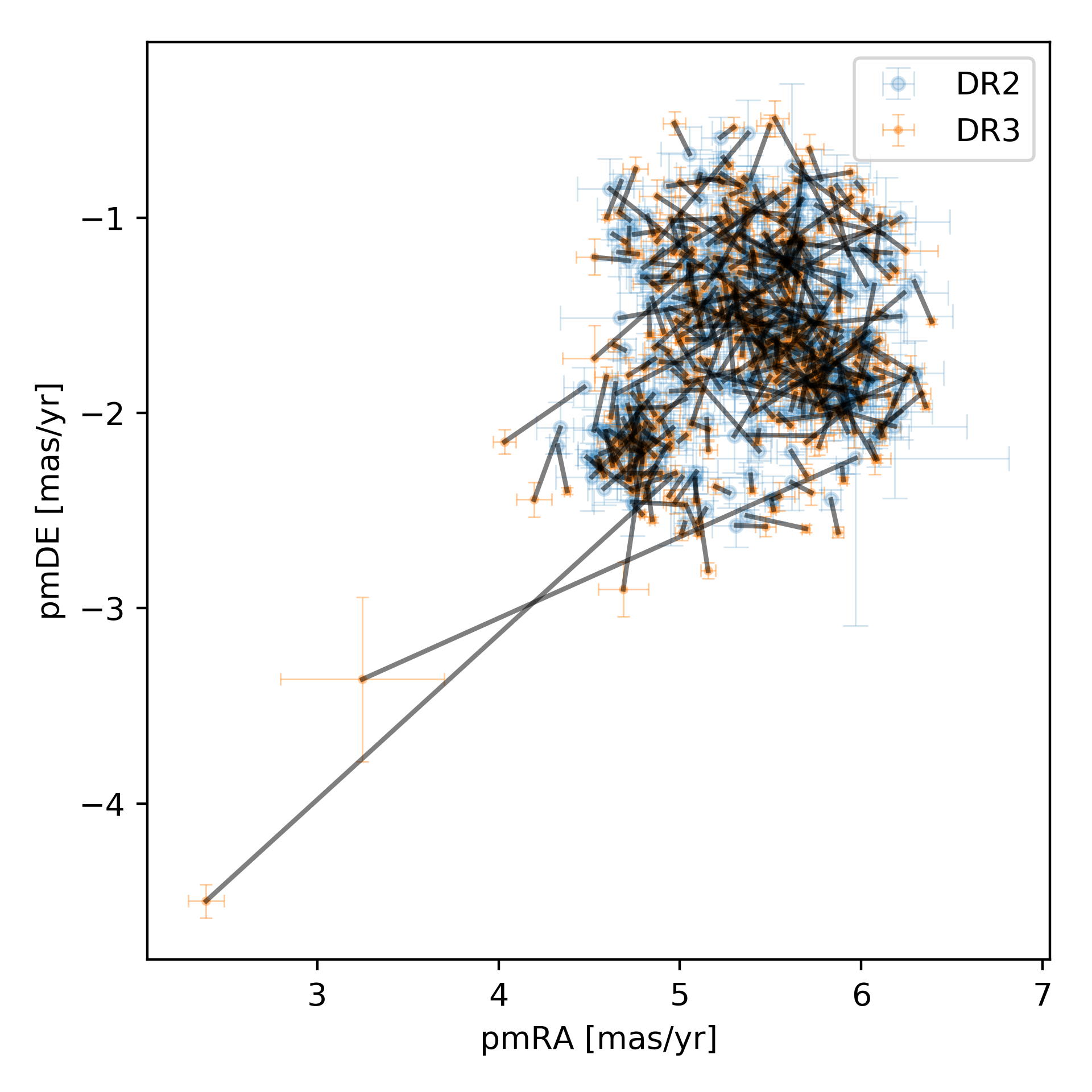}} \\
	 \end{tabular}
   \caption{Comparison of the Agg53 (RSG~7, RSG~8) cluster members determined by CG20 based on the DR2 and the EDR3 data; duplicates were filtered out. Top-left: the parallaxes of the clusters members. The red line represents the one-to-one relation. Top-right: the histogram of parallaxes. Bottom-left: the $\mu_{\alpha}$ component of the proper motion. The red line represents the one-to-one relation. Bottom-right: the proper motion diagram. The shifts of the positions of points is indicated by the black lines.}
   \label{edr3}
\end{figure*}

\subsection{Gulliver~56 and UBC~73}

Once again, there are no cluster parameters available for these clusters. Gulliver~56 has already been recognised in CG18, while UBC~73 is another newly discovered cluster. They share 50 out of 104 stars, with UBC~73 containing no star which has not been previously assigned to Gulliver~56.

The clusters occupy the same (extremely small) volume of the phase space (Fig.~\ref{phasespace_4}), although they do slightly differ in the proper motion component $\mu_{\alpha}$. The histogram in the top panel of Fig.~\ref{phasespace_4} shows that the previously derived parallaxes differ from what a fit of a double-Gaussian function predicts. However, the histogram fitting procedure may be inaccurate because of the small number of cluster members.

\subsection{RSG~7 and RSG~8}

The RSG clusters were discovered by \citet{Roser2016} who estimated $\log t = 8.3$ for RSG~7 and $\log t = 8.5$ for RSG~8. The reddening value was found to be quite low, $E(B-V)<0.1$ mag. No additional sources are available for the cluster parameters of these objects.

From the total of 404 stars, 143 are shared among the two clusters (117 stars have summed probabilities larger than $1.0$). CG18 lists parallaxes $\varpi = 2.337 \pm 0.008$ mas for RSG~7, and $\varpi = 2.210 \pm 0.012$ mas for RSG~8. They are very well distinguished in the parallax distribution (Fig.~\ref{phasespace_5}, top panel). However, we find that at least one of the clusters has a parallax ($\varpi = 2.297 \pm 0.012$ mas and $\varpi = 1.984 \pm 0.010$ mas) which does not coincide with the values provided by CG18.

In the sky, they overlap and are difficult to distinguish from each other (Fig.~\ref{phasespace_5}, middle-left panel). More interesting is the proper motion space (Fig.~\ref{phasespace_5}, middle-right panel), where we can see at least two concentrations of stars -- a similar structure can be seen in the proper motion diagram of the tidally disrupted cluster Coma Ber in \citet{Tang2019}. We find that the two concentrations are not represented well by the determined membership probabilities. However, they also do not correspond to the parallax distribution well.

Looking at the colour-magnitude diagrams (Fig.~\ref{phasespace_5}, bottom panels), the two clusters are almost indistinguishable. In \citet{Soubiran2019}, this aggregate was paired with another cluster, ASCC 127. It should be noted that the RSG clusters significantly differ from ASCC 127 in the colour-magnitude diagram. In our analysis, this cluster was most likely excluded due to the differences in the occupied volumes of the phase space.

\subsection{Kronberger~1 and Stock~8}

Three subgroups can be identified in the aggregate composed of five clusters. The overall distribution of parallaxes of this aggregate (Fig.~\ref{phasespace_6}, top-left panel) is even more complicated than in the case of Agg02. However, two clusters (Fig.~\ref{phasespace_6}, top-right panel) seem to deviate from the other ones.

COIN-Gaia~40 is a new cluster which does not share much in the phase space with the other clusters (Fig.~\ref{phasespace_6}, middle panels). It has been assigned 404 stars of which only 12 are shared with the rest of the aggregate.

Gulliver~53 and NGC~1893 share only four stars with each other (none with $p_{\textrm{sum}}>1.0$). They have similar central coordinates but differ in parallaxes and proper motion.

Information about Kronberger~1 has been presented by \citet{Sujatha2016}, who derived its age ($\log t = 7.65$) and the reddening ($E(B-V) = 0.45$ mag) in the line of sight. Cluster parameters are also available from SAI Open Clusters Catalog based on 2MASS photometry ($\log t = 8.10$, $E(B-V) = 0.43$ mag). However, the derived distance (800 pc) is significantly different from the one derived using Gaia photometry ($>2000$ pc).

Stock~8 is a young open cluster that has been known since \citet{Dias2002} determined its parameters ($\log t = 6.300$, $E(B-V) = 0.400$ mag). \citet{Kharchenko2013} present slightly different results, with a higher cluster age, $\log t = 7.050$, and reddening, $E(B-V) = 0.604$ mag. None of the works provided an estimate of the metallicity.

Kronberger~1 and Stock~8 can be somewhat distinguished in the coordinates space but CG18 give almost identical parallaxes. They share most of the members assigned to Kronberger~1, 37 out of 778 stars available in this sub-group (29 stars with $p_{\textrm{sum}}>1.0$). Their relative distance is estimated to be less than 20 pc. The mismatch in the derived cluster ages suggests that these clusters are quite different objects. However, it is possible that the age estimation is inaccurate for at least one of the clusters -- even for the much older Kronberger~1, the spread in the colour-magnitude diagrams is fairly large (Fig.~\ref{phasespace_6}, bottom panels).

Figure~\ref{phasespace_6} shows that the distribution of parallaxes does not coincide with the published values of parallaxes. Instead, our analysis puts the two clusters at much larger distances from each other, which would better explain the discrepancy between the cluster ages.

\section{Comparison with Gaia EDR3 and 2MASS}\label{section:4}

At the beginning of December 2020, Gaia Early Data Release 3 \citep[Gaia EDR3,][]{GaiaEDR3a,GaiaEDR3b} became available. $G$-magnitudes are now systematically shifted towards higher values (less bright). The values of parallaxes and proper motions have slightly changed, as well. However, they should still represent the same statistical result, in a given subset of stars, as before.

The lack of reliable radial velocity data means that it will be difficult to establish the binary fraction in a given subset of stars. The binaries may affect the search for the clusters -- we should expect that at least some of the members will occupy a different volume of the phase space than previously assumed.

Clearly, the membership probabilities have to be assessed anew. However, it would be interesting to look at the stars from CG20 and compare the situation between Gaia DR2 and Gaia EDR3. We searched for the cluster members of RSG~7 and RSG~8 and display the results in Fig.~\ref{edr3}. We can see that the parallaxes are systematically shifted from the one-to-one relation towards higher values than previously believed. On the other hand, the histogram of parallaxes did not change too much -- only the systematic shift towards higher values stands out. We see no systematic changes in the proper motion of the stars.

There are only a few outliers in the subplots. Therefore, we should expect that only a negligible number of cluster members is going to change (the most striking cases are seen in the proper motion diagram). Statistically, RSG~7 and RSG~8 now look similar to what we have seen in DR2. We conclude that the results of the analysis presented in this work should also remain valid in the new data release.

It is possible to search for the Gaia sources in other photometric catalogues based on the coordinates of the stars. We carried out this exercise for the 2MASS catalogue, where we searched for the stars of the six analysed aggregates. As we can see in the bottom-right panel of Fig.~\ref{phasespace_1}, Agg07 still shows a clear distinction in this photometric bands. Two populations also appear in the colour-magnitude diagram of Agg53 (Fig.~\ref{phasespace_6}). For the other aggregates, the situation is much more complicated -- at least for Agg32 and Agg42 (Fig.~\ref{phasespace_2} and Fig.~\ref{phasespace_4}, respectively), there seems to be no distinction between the two constituent clusters.

\section{Conclusions}\label{section:5}

Based on our analysis of the aggregates (derived from the stars which were assigned to multiple clusters), we conclude that the automatic procedure used in CG18 and CG20 failed to correctly determine parallaxes of some of the clusters listed in Table~\ref{table-1}:

\begin{enumerate}
    \item Alessi~5 and BH~99: The central parallaxes from the source work were correctly reproduced by analysing the distribution of parallaxes. We propose that the match results from the fact that the clusters are only partially overlapping in the phase space. Another important contributing factor is that only one of the stars is shared among them. Assuming that their relative distance is about 50 pc, it is interesting to note that the metallicity and the extinction (taken from the literature) differ for these clusters -- this can be seen as a population difference in the colour-magnitude diagram of this aggregate.
    \item FSR~0686 and UBC~55: Our fitting method gives different parallaxes than presented in the literature -- this is most likely due to the large amount of duplicate stars in this aggregate. However, the total number of members is fairly low, which seriously affects the usefulness of the method we used. The distribution can be well fitted by a single Gaussian function which would yield a value similar to those from CG20.
    \item Gulliver~6 and UBC~17b: In this case, the cited parallaxes differ only a little from what we see in the histogram. The shape of the parallax distribution is slightly asymmetrical. The two objects are also difficult to distinguish from each other in the coordinates space. We suggest that the aggregate could be somehow physically distorted, producing a significant deviation from normal distribution in the parallax space.
    \item Gulliver~56 and UBC~73: The fitting method gives parallaxes that are quite different from those presented in the source work. However, this case is very similar to Agg32 -- many stars are shared and the total number of stars is low.
    \item RSG~7 and RSG~8: This aggregate is very interesting. Two populations are apparent in the parallax space and the proper motion space, but they do not seem to correspond to each other. On the other hand, there seems to be no evidence in the coordinate space or the colour-magnitude diagram that this aggregate is composed of multiple clusters. Based on our analysis, we propose an update to the parallax values presented by CG18.
    \item Kronberger~1 and Stock~8: Most of the stars assigned to Kronberger~1 are also present in Stock~8. Our analysis of parallaxes puts the clusters at a larger relative distance than previously derived. This would better explain the difference between the ages found in the literature.
\end{enumerate}

Most of the cluster parallaxes listed in CG20 were derived using a very robust method. However, a problem with deriving the parallaxes seems to occur when at least two clusters overlap in a small volume of the phase space and when they share most of the members. This results from ignoring a possibility of encountering more complex aggregates of clusters in the phase space. In such cases, an identification of such aggregates followed by a separate analysis of the parallaxes is required.

We propose that the membership probabilities of stars in cluster aggregates, which seem to occur less frequently than single clusters, should be calculated anew. Most importantly, the probability of a star being a member of at least one of the assigned clusters (the summed probability) should be restricted by the condition $p_{\textrm{sum}} \leq 1.0$.

\begin{acknowledgements}
This paper is dedicated to Harald P{\"o}hnl who died during its preparation.
It was supported by an internal grant of the Masaryk University (MUNI/IGA/1530/2020).
EP acknowledges support by
the Erasmus+ programme of the European Union under grant number
2020-1-CZ01-KA203-078200.
This work made use of data from the European Space Agency (ESA) mission {\it Gaia} (\url{https://www.cosmos.esa.int/gaia}), 
processed by the {\it Gaia} Data Processing and Analysis Consortium (DPAC, \url{https://www.cosmos.esa.int/web/gaia/dpac/consortium}). 
Funding for the DPAC has been provided by national institutions, in particular the institutions participating in the {\it Gaia} 
Multilateral Agreement. This research has made use of the SIMBAD database,
operated at CDS, Strasbourg, France.
This research has made use of the WEBDA database, operated at the Department of Theoretical Physics and Astrophysics of the Masaryk University.
\end{acknowledgements}

\bibliographystyle{aa} 
\bibliography{aanda}

\begin{thebibliography}{19}
\expandafter\ifx\csname natexlab\endcsname\relax\def\natexlab#1{#1}\fi

\bibitem[{{Bossini} {et~al.}(2019){Bossini}, {Vallenari}, {Bragaglia},
  {Cantat-Gaudin}, {Sordo}, {Balaguer-N{\'u}{\~n}ez}, {Jordi}, {Moitinho},
  {Soubiran}, {Casamiquela}, {Carrera}, \& {Heiter}}]{Bossini2019}
{Bossini}, D., {Vallenari}, A., {Bragaglia}, A., {et~al.} 2019, \aap, 623, A108

\bibitem[{{Cantat-Gaudin} \& {Anders}(2020)}]{Cantat2020}
{Cantat-Gaudin}, T. \& {Anders}, F. 2020, \aap, 633, A99

\bibitem[{{Cantat-Gaudin} {et~al.}(2018){Cantat-Gaudin}, {Jordi}, {Vallenari},
  {Bragaglia}, {Balaguer-N{\'u}{\~n}ez}, {Soubiran}, {Bossini}, {Moitinho},
  {Castro-Ginard}, {Krone-Martins}, {Casamiquela}, {Sordo}, \&
  {Carrera}}]{Cantat2018}
{Cantat-Gaudin}, T., {Jordi}, C., {Vallenari}, A., {et~al.} 2018, \aap, 618,
  A93

\bibitem[{{Dehnen} \& {Binney}(1998)}]{Dehnen1998}
{Dehnen}, W. \& {Binney}, J.~J. 1998, \mnras, 298, 387

\bibitem[{{Dias} {et~al.}(2002){Dias}, {Alessi}, {Moitinho}, \&
  {L{\'e}pine}}]{Dias2002}
{Dias}, W.~S., {Alessi}, B.~S., {Moitinho}, A., \& {L{\'e}pine}, J.~R.~D. 2002,
  \aap, 389, 871

\bibitem[{{Gaia Collaboration} {et~al.}(2018){Gaia Collaboration}, {Brown},
  {Vallenari}, {Prusti}, {de Bruijne}, {Babusiaux}, {Bailer-Jones}, {Biermann},
  {Evans}, {Eyer}, {Jansen}, {Jordi}, {Klioner}, {Lammers}, {Lindegren},
  {Luri}, {Mignard}, {Panem}, {Pourbaix}, {Randich}, {Sartoretti}, {Siddiqui},
  {Soubiran}, {van Leeuwen}, {Walton}, {Arenou}, {Bastian}, {Cropper},
  {Drimmel}, {Katz}, {Lattanzi}, {Bakker}, {Cacciari}, {Casta{\~n}eda},
  {Chaoul}, {Cheek}, {De Angeli}, {Fabricius}, {Guerra}, {Holl}, {Masana},
  {Messineo}, {Mowlavi}, {Nienartowicz}, {Panuzzo}, {Portell}, {Riello},
  {Seabroke}, {Tanga}, {Th{\'e}venin}, {Gracia-Abril}, {Comoretto},
  {Garcia-Reinaldos}, {Teyssier}, {Altmann}, {Andrae}, {Audard},
  {Bellas-Velidis}, {Benson}, {Berthier}, {Blomme}, {Burgess}, {Busso},
  {Carry}, {Cellino}, {Clementini}, {Clotet}, {Creevey}, {Davidson}, {De
  Ridder}, {Delchambre}, {Dell'Oro}, {Ducourant},
  {Fern{\'a}ndez-Hern{\'a}ndez}, {Fouesneau}, {Fr{\'e}mat}, {Galluccio},
  {Garc{\'\i}a-Torres}, {Gonz{\'a}lez-N{\'u}{\~n}ez}, {Gonz{\'a}lez-Vidal},
  {Gosset}, {Guy}, {Halbwachs}, {Hambly}, {Harrison}, {Hern{\'a}ndez},
  {Hestroffer}, {Hodgkin}, {Hutton}, {Jasniewicz}, {Jean-Antoine-Piccolo},
  {Jordan}, {Korn}, {Krone-Martins}, {Lanzafame}, {Lebzelter}, {L{\"o}ffler},
  {Manteiga}, {Marrese}, {Mart{\'\i}n-Fleitas}, {Moitinho}, {Mora}, {Muinonen},
  {Osinde}, {Pancino}, {Pauwels}, {Petit}, {Recio-Blanco}, {Richards},
  {Rimoldini}, {Robin}, {Sarro}, {Siopis}, {Smith}, {Sozzetti}, {S{\"u}veges},
  {Torra}, {van Reeven}, {Abbas}, {Abreu Aramburu}, {Accart}, {Aerts},
  {Altavilla}, {{\'A}lvarez}, {Alvarez}, {Alves}, {Anderson}, {Andrei},
  {Anglada Varela}, {Antiche}, {Antoja}, {Arcay}, {Astraatmadja}, {Bach},
  {Baker}, {Balaguer-N{\'u}{\~n}ez}, {Balm}, {Barache}, {Barata}, {Barbato},
  {Barblan}, {Barklem}, {Barrado}, {Barros}, {Barstow}, {Bartholom{\'e}
  Mu{\~n}oz}, {Bassilana}, {Becciani}, {Bellazzini}, {Berihuete}, {Bertone},
  {Bianchi}, {Bienaym{\'e}}, {Blanco-Cuaresma}, {Boch}, {Boeche}, {Bombrun},
  {Borrachero}, {Bossini}, {Bouquillon}, {Bourda}, {Bragaglia}, {Bramante},
  {Breddels}, {Bressan}, {Brouillet}, {Br{\"u}semeister}, {Brugaletta},
  {Bucciarelli}, {Burlacu}, {Busonero}, {Butkevich}, {Buzzi}, {Caffau},
  {Cancelliere}, {Cannizzaro}, {Cantat-Gaudin}, {Carballo}, {Carlucci},
  {Carrasco}, {Casamiquela}, {Castellani}, {Castro-Ginard}, {Charlot},
  {Chemin}, {Chiavassa}, {Cocozza}, {Costigan}, {Cowell}, {Crifo}, {Crosta},
  {Crowley}, {Cuypers}, {Dafonte}, {Damerdji}, {Dapergolas}, {David}, {David},
  {de Laverny}, {De Luise}, {De March}, {de Martino}, {de Souza}, {de Torres},
  {Debosscher}, {del Pozo}, {Delbo}, {Delgado}, {Delgado}, {Di Matteo},
  {Diakite}, {Diener}, {Distefano}, {Dolding}, {Drazinos}, {Dur{\'a}n},
  {Edvardsson}, {Enke}, {Eriksson}, {Esquej}, {Eynard Bontemps}, {Fabre},
  {Fabrizio}, {Faigler}, {Falc{\~a}o}, {Farr{\`a}s Casas}, {Federici},
  {Fedorets}, {Fernique}, {Figueras}, {Filippi}, {Findeisen}, {Fonti},
  {Fraile}, {Fraser}, {Fr{\'e}zouls}, {Gai}, {Galleti}, {Garabato},
  {Garc{\'\i}a-Sedano}, {Garofalo}, {Garralda}, {Gavel}, {Gavras}, {Gerssen},
  {Geyer}, {Giacobbe}, {Gilmore}, {Girona}, {Giuffrida}, {Glass}, {Gomes},
  {Granvik}, {Gueguen}, {Guerrier}, {Guiraud}, {Guti{\'e}rrez-S{\'a}nchez},
  {Haigron}, {Hatzidimitriou}, {Hauser}, {Haywood}, {Heiter}, {Helmi}, {Heu},
  {Hilger}, {Hobbs}, {Hofmann}, {Holland}, {Huckle}, {Hypki}, {Icardi},
  {Jan{\ss}en}, {Jevardat de Fombelle}, {Jonker}, {Juh{\'a}sz}, {Julbe},
  {Karampelas}, {Kewley}, {Klar}, {Kochoska}, {Kohley}, {Kolenberg},
  {Kontizas}, {Kontizas}, {Koposov}, {Kordopatis}, {Kostrzewa-Rutkowska},
  {Koubsky}, {Lambert}, {Lanza}, {Lasne}, {Lavigne}, {Le Fustec}, {Le
  Poncin-Lafitte}, {Lebreton}, {Leccia}, {Leclerc}, {Lecoeur-Taibi},
  {Lenhardt}, {Leroux}, {Liao}, {Licata}, {Lindstr{\o}m}, {Lister}, {Livanou},
  {Lobel}, {L{\'o}pez}, {Managau}, {Mann}, {Mantelet}, {Marchal}, {Marchant},
  {Marconi}, {Marinoni}, {Marschalk{\'o}}, {Marshall}, {Martino}, {Marton},
  {Mary}, {Massari}, {Matijevi{\v{c}}}, {Mazeh}, {McMillan}, {Messina},
  {Michalik}, {Millar}, {Molina}, {Molinaro}, {Moln{\'a}r}, {Montegriffo},
  {Mor}, {Morbidelli}, {Morel}, {Morris}, {Mulone}, {Muraveva}, {Musella},
  {Nelemans}, {Nicastro}, {Noval}, {O'Mullane}, {Ord{\'e}novic},
  {Ord{\'o}{\~n}ez-Blanco}, {Osborne}, {Pagani}, {Pagano}, {Pailler},
  {Palacin}, {Palaversa}, {Panahi}, {Pawlak}, {Piersimoni}, {Pineau}, {Plachy},
  {Plum}, {Poggio}, {Poujoulet}, {Pr{\v{s}}a}, {Pulone}, {Racero}, {Ragaini},
  {Rambaux}, {Ramos-Lerate}, {Regibo}, {Reyl{\'e}}, {Riclet}, {Ripepi}, {Riva},
  {Rivard}, {Rixon}, {Roegiers}, {Roelens}, {Romero-G{\'o}mez}, {Rowell},
  {Royer}, {Ruiz-Dern}, {Sadowski}, {Sagrist{\`a} Sell{\'e}s}, {Sahlmann},
  {Salgado}, {Salguero}, {Sanna}, {Santana-Ros}, {Sarasso}, {Savietto},
  {Schultheis}, {Sciacca}, {Segol}, {Segovia}, {S{\'e}gransan}, {Shih},
  {Siltala}, {Silva}, {Smart}, {Smith}, {Solano}, {Solitro}, {Sordo}, {Soria
  Nieto}, {Souchay}, {Spagna}, {Spoto}, {Stampa}, {Steele},
  {Steidelm{\"u}ller}, {Stephenson}, {Stoev}, {Suess}, {Surdej}, {Szabados},
  {Szegedi-Elek}, {Tapiador}, {Taris}, {Tauran}, {Taylor}, {Teixeira},
  {Terrett}, {Teyssand ier}, {Thuillot}, {Titarenko}, {Torra Clotet}, {Turon},
  {Ulla}, {Utrilla}, {Uzzi}, {Vaillant}, {Valentini}, {Valette}, {van Elteren},
  {Van Hemelryck}, {van Leeuwen}, {Vaschetto}, {Vecchiato}, {Veljanoski},
  {Viala}, {Vicente}, {Vogt}, {von Essen}, {Voss}, {Votruba}, {Voutsinas},
  {Walmsley}, {Weiler}, {Wertz}, {Wevers}, {Wyrzykowski}, {Yoldas},
  {{\v{Z}}erjal}, {Ziaeepour}, {Zorec}, {Zschocke}, {Zucker}, {Zurbach}, \&
  {Zwitter}}]{GaiaDR2}
{Gaia Collaboration}, {Brown}, A.~G.~A., {Vallenari}, A., {et~al.} 2018, \aap,
  616, A1

\bibitem[{{Gaia Collaboration} {et~al.}(2020){Gaia Collaboration}, {Brown},
  {Vallenari}, {Prusti}, {de Bruijne}, {Babusiaux}, \& {Biermann}}]{GaiaEDR3a}
{Gaia Collaboration}, {Brown}, A.~G.~A., {Vallenari}, A., {et~al.} 2020, arXiv
  e-prints, arXiv:2012.01533

\bibitem[{{Goodwin}(1997)}]{Goodwin1997}
{Goodwin}, S.~P. 1997, \mnras, 284, 785

\bibitem[{{Kharchenko} {et~al.}(2013){Kharchenko}, {Piskunov}, {Schilbach},
  {R{\"o}ser}, \& {Scholz}}]{Kharchenko2013}
{Kharchenko}, N.~V., {Piskunov}, A.~E., {Schilbach}, E., {R{\"o}ser}, S., \&
  {Scholz}, R.~D. 2013, \aap, 558, A53

\bibitem[{{Krone-Martins} \& {Moitinho}(2014)}]{KroneMartins14}
{Krone-Martins}, A. \& {Moitinho}, A. 2014, \aap, 561, A57

\bibitem[{{Lindegren} {et~al.}(2020){Lindegren}, {Bastian}, {Biermann},
  {Bombrun}, {de Torres}, {Gerlach}, {Geyer}, {Hern{\'a}ndez}, {Hilger},
  {Hobbs}, {Klioner}, {Lammers}, {McMillan}, {Ramos-Lerate},
  {Steidelm{\"u}ller}, {Stephenson}, \& {van Leeuwen}}]{GaiaEDR3b}
{Lindegren}, L., {Bastian}, U., {Biermann}, M., {et~al.} 2020, arXiv e-prints,
  arXiv:2012.01742

\bibitem[{{Luri} {et~al.}(2018){Luri}, {Brown}, {Sarro}, {Arenou},
  {Bailer-Jones}, {Castro-Ginard}, {de Bruijne}, {Prusti}, {Babusiaux}, \&
  {Delgado}}]{Luri2018}
{Luri}, X., {Brown}, A.~G.~A., {Sarro}, L.~M., {et~al.} 2018, \aap, 616, A9

\bibitem[{{Priyatikanto} {et~al.}(2016){Priyatikanto}, {Kouwenhoven},
  {Arifyanto}, {Wulandari}, \& {Siregar}}]{Priyatikanto2016}
{Priyatikanto}, R., {Kouwenhoven}, M.~B.~N., {Arifyanto}, M.~I., {Wulandari},
  H.~R.~T., \& {Siregar}, S. 2016, \mnras, 457, 1339

\bibitem[{{R{\"o}ser} {et~al.}(2016){R{\"o}ser}, {Schilbach}, \&
  {Goldman}}]{Roser2016}
{R{\"o}ser}, S., {Schilbach}, E., \& {Goldman}, B. 2016, \aap, 595, A22

\bibitem[{{Soubiran} {et~al.}(2018){Soubiran}, {Cantat-Gaudin},
  {Romero-G{\'o}mez}, {Casamiquela}, {Jordi}, {Vallenari}, {Antoja},
  {Balaguer-N{\'u}{\~n}ez}, {Bossini}, {Bragaglia}, {Carrera}, {Castro-Ginard},
  {Figueras}, {Heiter}, {Katz}, {Krone-Martins}, {Le Campion}, {Moitinho}, \&
  {Sordo}}]{Soubiran2018}
{Soubiran}, C., {Cantat-Gaudin}, T., {Romero-G{\'o}mez}, M., {et~al.} 2018,
  \aap, 619, A155

\bibitem[{{Soubiran} {et~al.}(2019){Soubiran}, {Cantat-Gaudin},
  {Romero-G{\'o}mez}, {Casamiquela}, {Jordi}, {Vallenari}, {Antoja},
  {Balaguer-N{\'u}{\~n}ez}, {Bossini}, {Bragaglia}, {Carrera}, {Castro-Ginard},
  {Figueras}, {Heiter}, {Katz}, {Krone-Martins}, {Le Campion}, {Moitinho}, \&
  {Sordo}}]{Soubiran2019}
{Soubiran}, C., {Cantat-Gaudin}, T., {Romero-G{\'o}mez}, M., {et~al.} 2019,
  \aap, 623, C2

\bibitem[{{Sujatha} {et~al.}(2016){Sujatha}, {Krishna Kumar}, \&
  {Komala}}]{Sujatha2016}
{Sujatha}, S., {Krishna Kumar}, K., \& {Komala}, S. 2016, \actaa, 66, 333

\bibitem[{{Tang} {et~al.}(2019){Tang}, {Pang}, {Yuan}, {Chen}, {Hong},
  {Goldman}, {Just}, {Shukirgaliyev}, \& {Lin}}]{Tang2019}
{Tang}, S.-Y., {Pang}, X., {Yuan}, Z., {et~al.} 2019, \apj, 877, 12

\bibitem[{{van den Bergh}(1996)}]{vdBergh1996}
{van den Bergh}, S. 1996, \apjl, 471, L31

\end{thebibliography}

\begin{appendix}
\section{Appendix}\label{section:Appendix}

In this section, we display the coordinate space map, the proper motion diagram, the parallax distribution, and the colour-magnitude diagram of each of the six analysed aggregates. Included are all stars belonging to all clusters of the whole aggregates (except for the parallax space, where duplicates were excluded). Moreover, the complete list of all aggregates we found is presented in Table~\ref{table-A}.

\newpage

\begin{table*}
\caption{The complete list of aggregates of clusters.}
\label{table-A}
\centering
\begin{tabular}{c | l}
\hline\hline
Aggregate ID & List of clusters \\
\hline
    Agg01 & ASCC 105, UPK 82 \\
    Agg02 & ASCC 19, Gulliver 6, UBC 17a, UBC 17b \\
    Agg03 & ASCC 88, Gulliver 29 \\
    Agg04 & Alessi 2, UPK 312 \\
    Agg05 & Alessi 43, Collinder 197 \\
    Agg06 & Alessi 44, UBC 14 \\
    Agg07 & Alessi 5, BH 99 \\
    Agg08 & Alessi Teutsch 5, BDSB30 \\
    Agg09 & BH 121, IC 2948, Ruprecht 94 \\
    Agg10 & Barkhatova 1, Gulliver 30 \\
    Agg11 & Berkeley 58, NGC 7788, NGC 7790 \\
    Agg12 & Berkeley 62, COIN-Gaia 29 \\
    Agg13 & Berkeley 81, NGC 6735 \\
    Agg14 & Biurakan 2, FSR 0198, NGC 6871, Teutsch 8 \\
    Agg15 & COIN-Gaia 16, COIN-Gaia 17 \\
    Agg16 & COIN-Gaia 24, NGC 2168 \\
    Agg17 & COIN-Gaia 40, Gulliver 53, Kronberger 1, NGC 1893, Stock 8 \\
    Agg18 & COIN-Gaia 5, COIN-Gaia 6 \\
    Agg19 & Collinder 106, Collinder 107, NGC 2244 \\
    Agg20 & Collinder 135, UBC 7 \\
    Agg21 & Collinder 220, IC 2581 \\
    Agg22 & Collinder 272, NGC 5168 \\
    Agg23 & Czernik 20, NGC 1857 \\
    Agg24 & Czernik 31, NGC 2421 \\
    Agg25 & Czernik 39, NGC 6755, NGC 6756 \\
    Agg26 & Danks 1, Danks 2 \\
    Agg27 & Dias 1, King 16 \\
    Agg28 & FSR 0306, NGC 7086 \\
    Agg29 & FSR 0448, FSR 0451 \\
    Agg30 & FSR 0534, UBC 39 \\
    Agg31 & FSR 0542, SAI 14 \\
    Agg32 & FSR 0686, UBC 55 \\
    Agg33 & FSR 0826, Teutsch 10 \\
    Agg34 & FSR 1207, NGC 2345 \\
    Agg35 & FSR 1315, NGC 2453 \\
    Agg36 & Feibelman 1, Gulliver 23 \\
    Agg37 & Gulliver 12, NGC 4103 \\
    Agg38 & Gulliver 15, NGC 6561 \\
    Agg39 & Gulliver 16, NGC 581 \\
    Agg40 & Gulliver 3, Gulliver 4 \\
    Agg41 & Gulliver 52, Trumpler 15 \\
    Agg42 & Gulliver 56, UBC 73 \\
    Agg43 & Haffner 18, Haffner 19 \\
    Agg44 & Hogg 10, NGC 3572 \\
    Agg45 & Hogg 17, NGC 5617, Pismis 19, Trumpler 22 \\
    Agg46 & Kharchenko 1, Koposov 63 \\
    Agg47 & King 14, NGC 146 \\
    Agg48 & NGC 2194, Skiff J0614+12.9 \\
    Agg49 & NGC 2318, Ruprecht 8 \\
    Agg50 & NGC 436, NGC 457 \\
    Agg51 & NGC 5269, SAI 118 \\
    Agg52 & NGC 659, NGC 663 \\
    Agg53 & RSG 7, RSG 8 \\
    Agg54 & Ruprecht 100, Ruprecht 101 \\
    Agg55 & Ruprecht 43, Ruprecht 44 \\
    Agg56 & SAI 4, Stock 20 \\
    Agg57 & Teutsch 11, Teutsch 12 \\
    Agg58 & UBC 10a, UPK 169 \\
    Agg59 & UPK 219, UPK 220 \\
    Agg60 & UPK 429, UPK 431 \\
\hline
\end{tabular}
\end{table*}

\newpage

\begin{figure*}[t]
   \centering
	 \begin{tabular}{ll}
   \multicolumn{2}{c}{\subfloat{\includegraphics[scale=0.5]{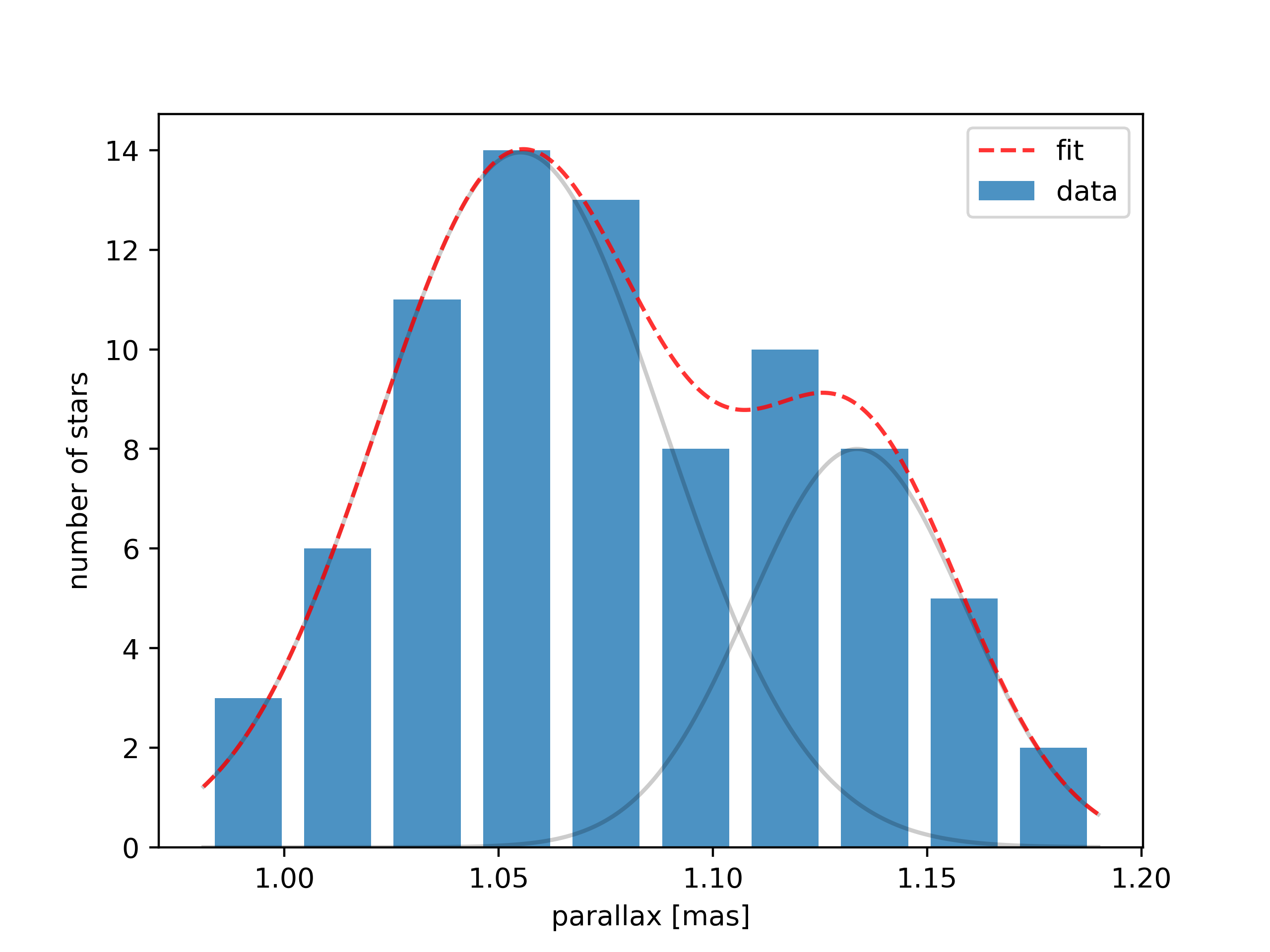}}} \\ 
   \subfloat{\includegraphics[scale=0.5]{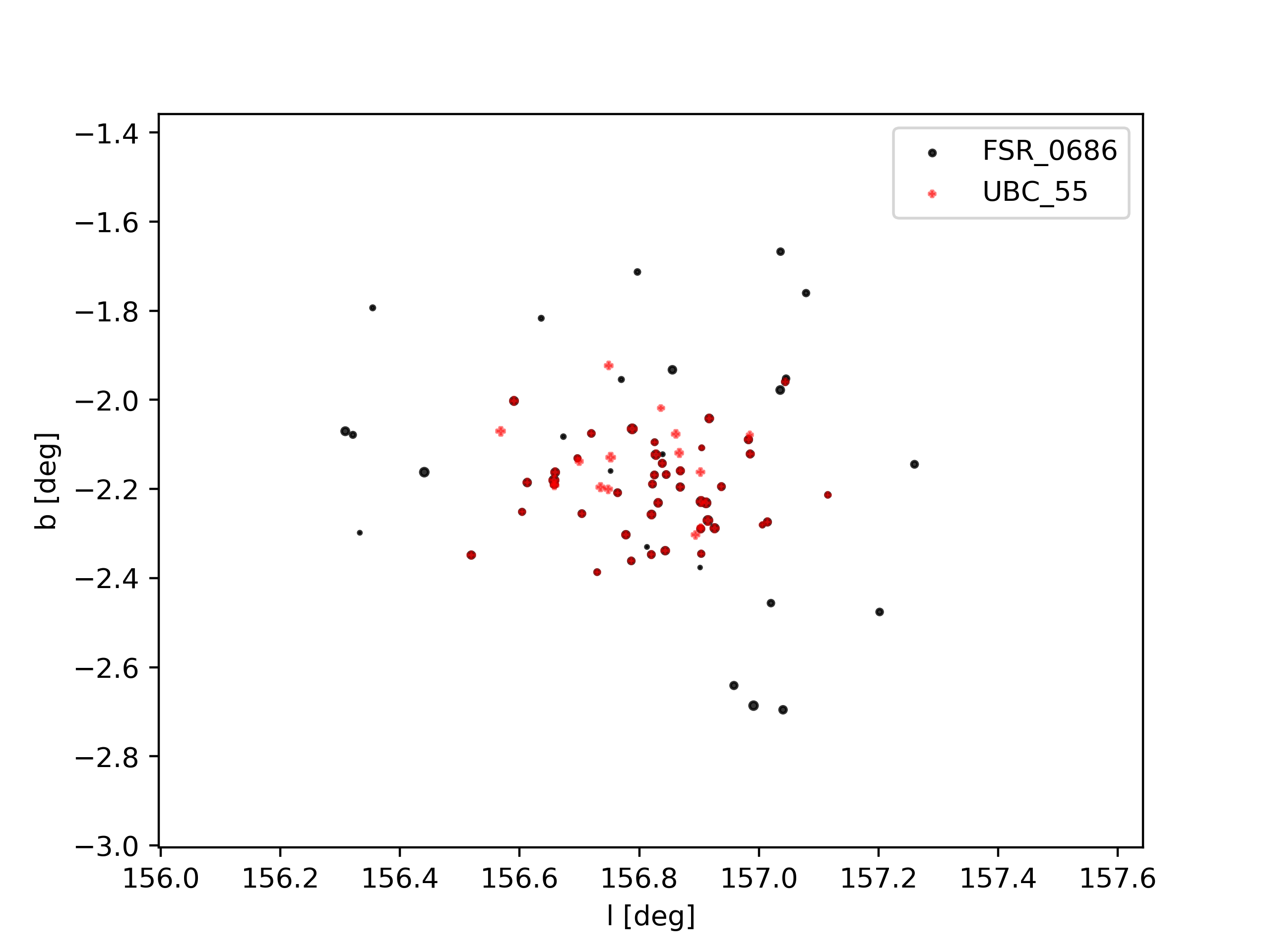}} \quad &
	 \subfloat{\includegraphics[scale=0.5]{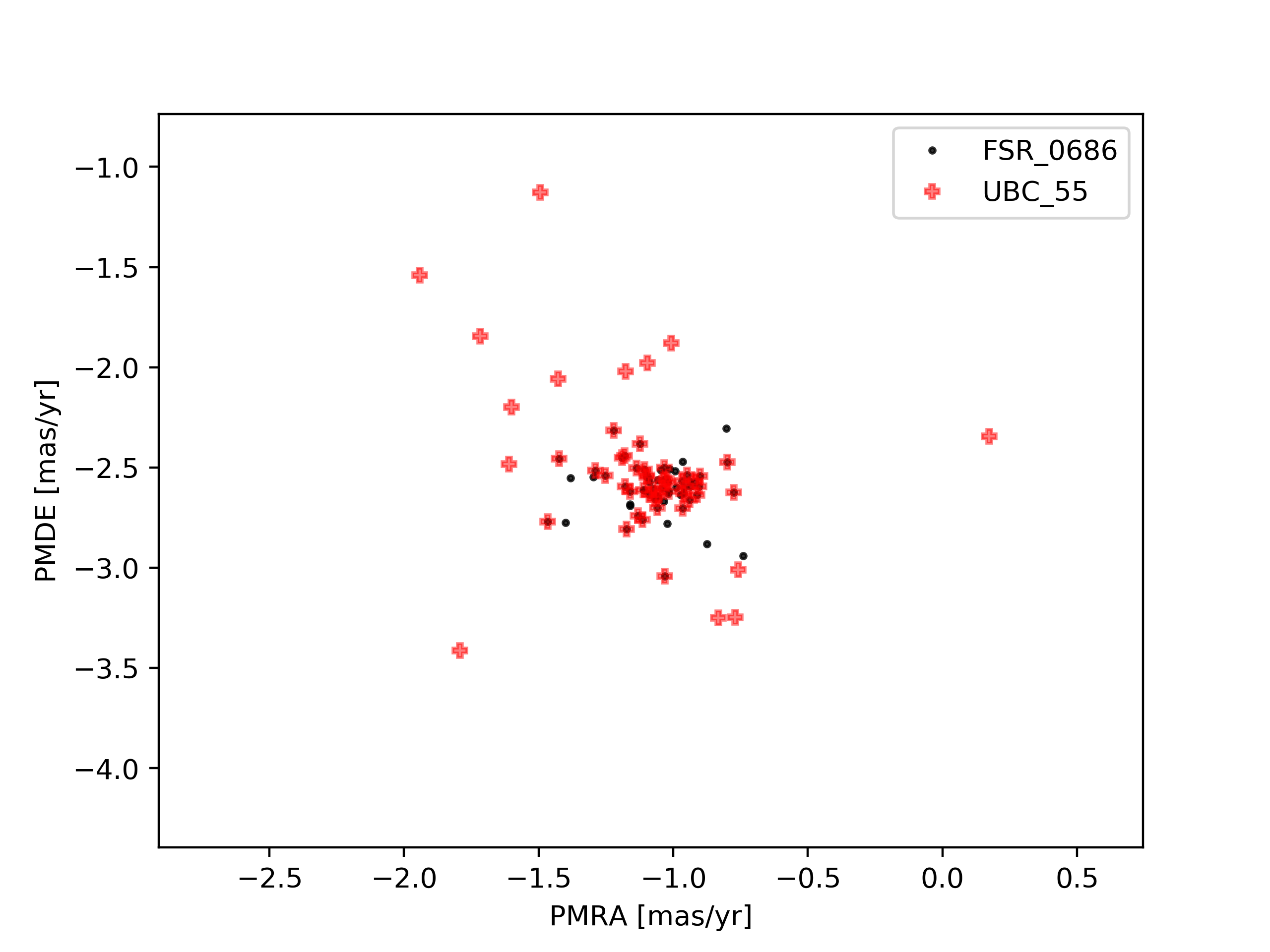}} \\
   \subfloat{\includegraphics[scale=0.5]{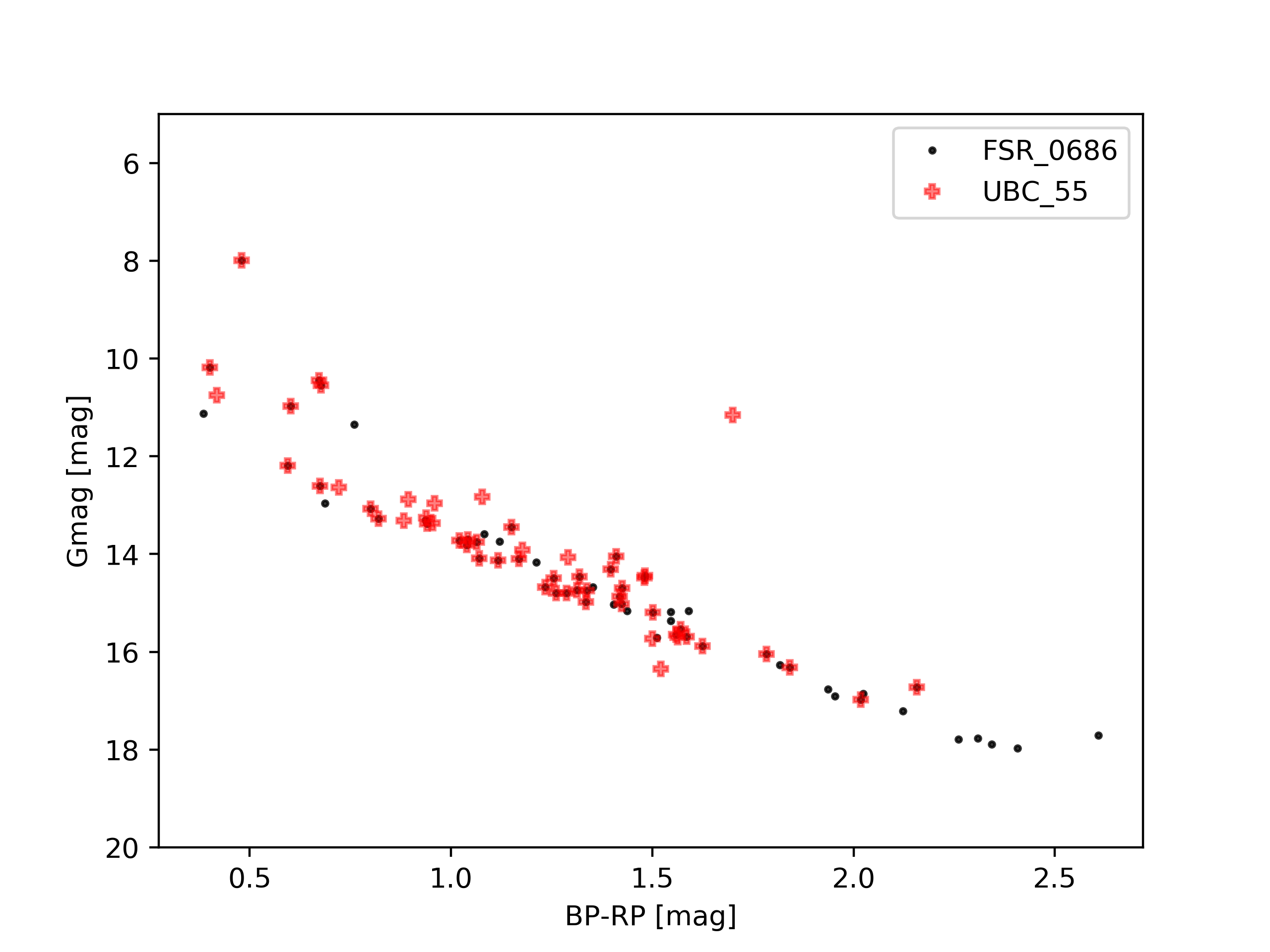}} \quad & 
   \subfloat{\includegraphics[scale=0.5]{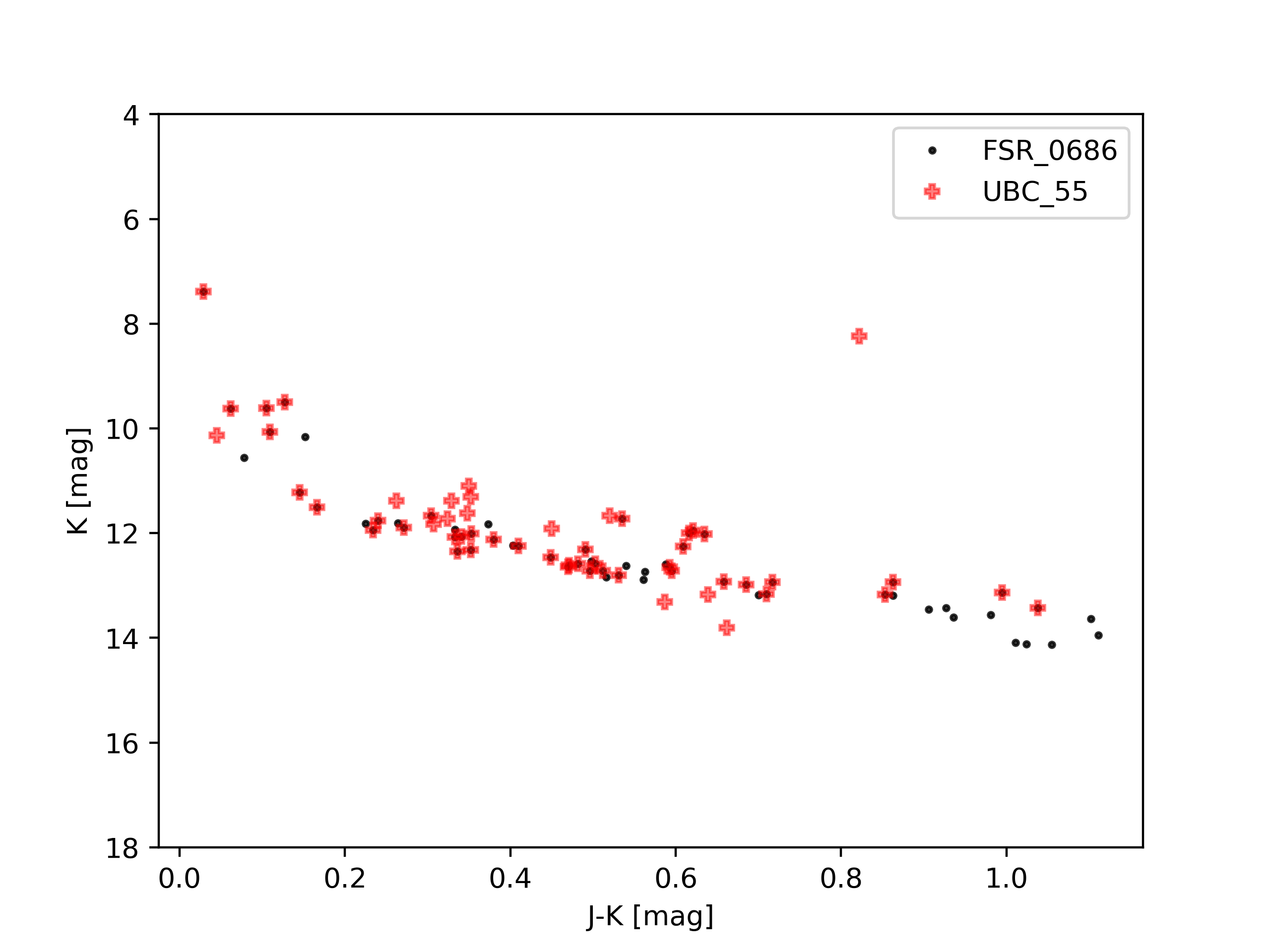}} \\
	 \end{tabular}
   \caption{Different slices through the phase space of Agg32 (FSR~0686, UBC~55), together with the complete colour-magnitude diagrams (based on Gaia and 2MASS photometry). Top: the histogram of parallaxes, excluding the duplicate cluster members. The best fit was achieved with the following double-Gaussian function parameters: $\varpi_1 = 1.055 \pm 0.005$ mas, $\sigma_1 = 0.033 \pm 0.004$ mas, $\varpi_2 = 1.134 \pm 0.007$ mas, $\sigma_2 = 0.025 \pm 0.005$ mas. Middle-left: coordinates of the stars in Agg32. Size of the points indicates values of the observed magnitude $G$. Middle-right: proper motion diagram of Agg32. Bottom-left: colour-magnitude diagram of Agg32, based on the observed values of $B_P-R_P$ and $G$). Bottom-right: colour-magnitude diagram of Agg32 based on 2MASS photometry. The stars were located using the coordinates in Gaia database.}
   \label{phasespace_2}
\end{figure*}

\newpage

\begin{figure*}[t]
   \centering
	 \begin{tabular}{ll}
   \subfloat{\includegraphics[scale=0.5]{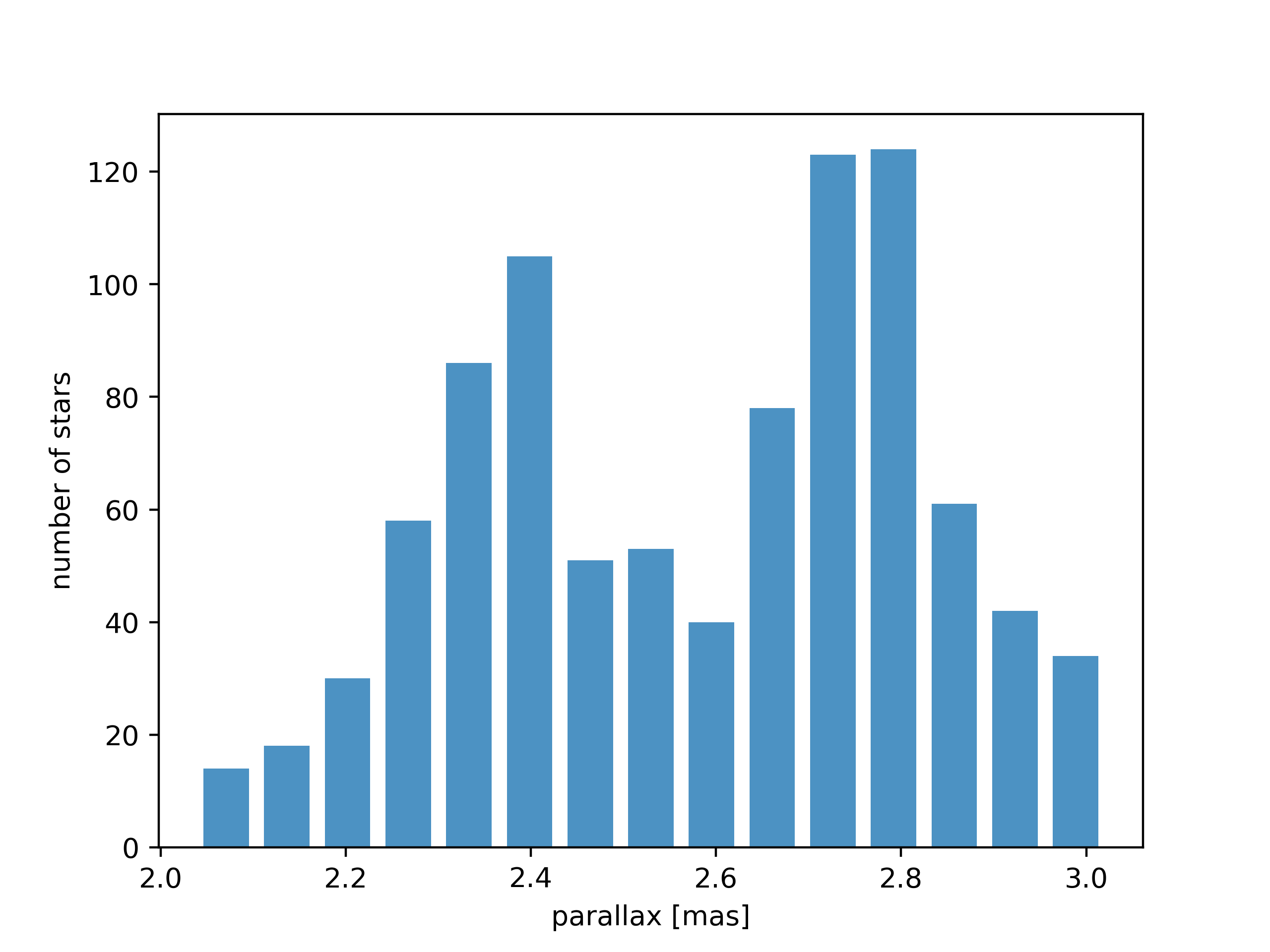}} \quad & 
   \subfloat{\includegraphics[scale=0.5]{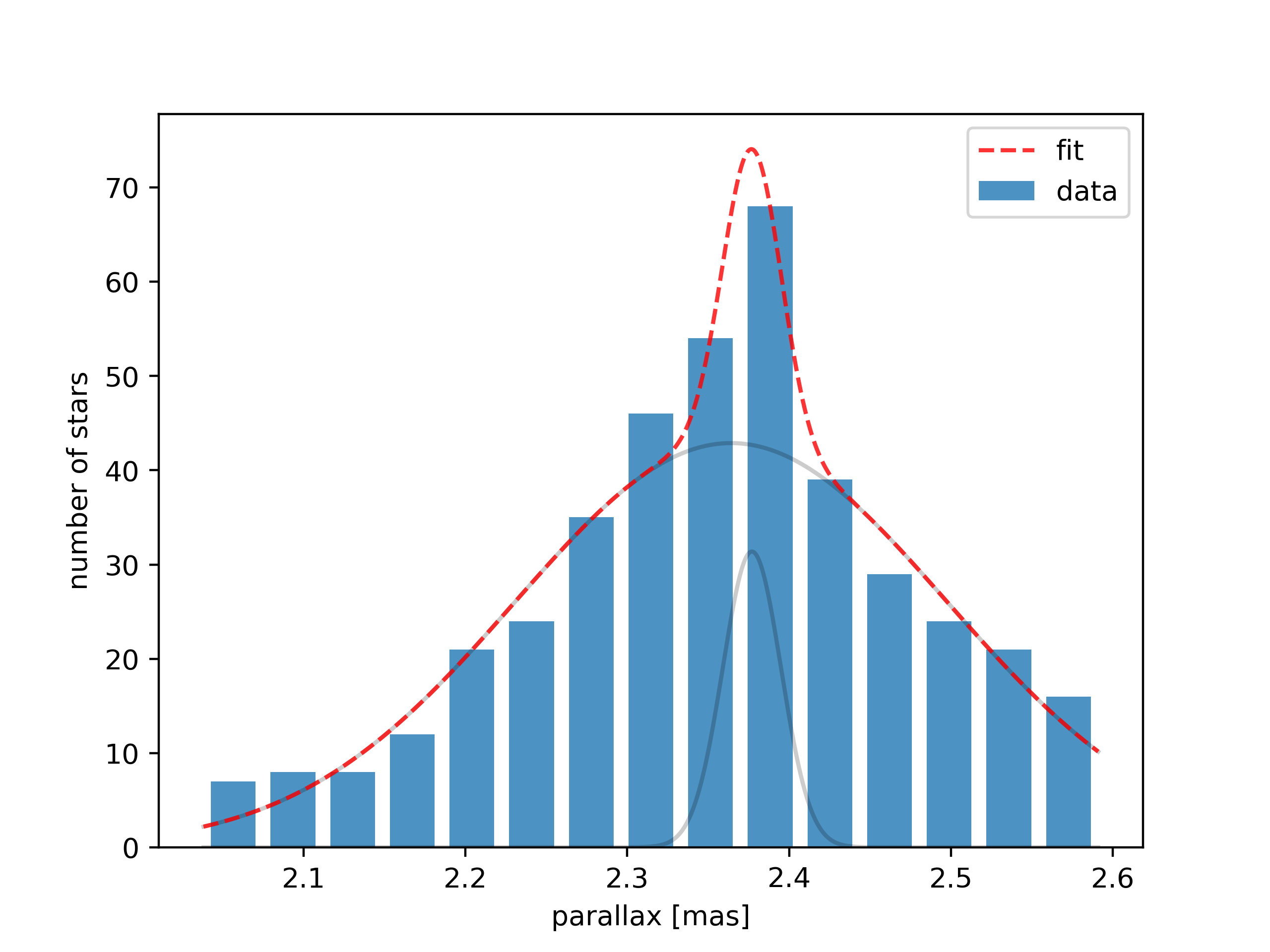}} \\
   \subfloat{\includegraphics[scale=0.5]{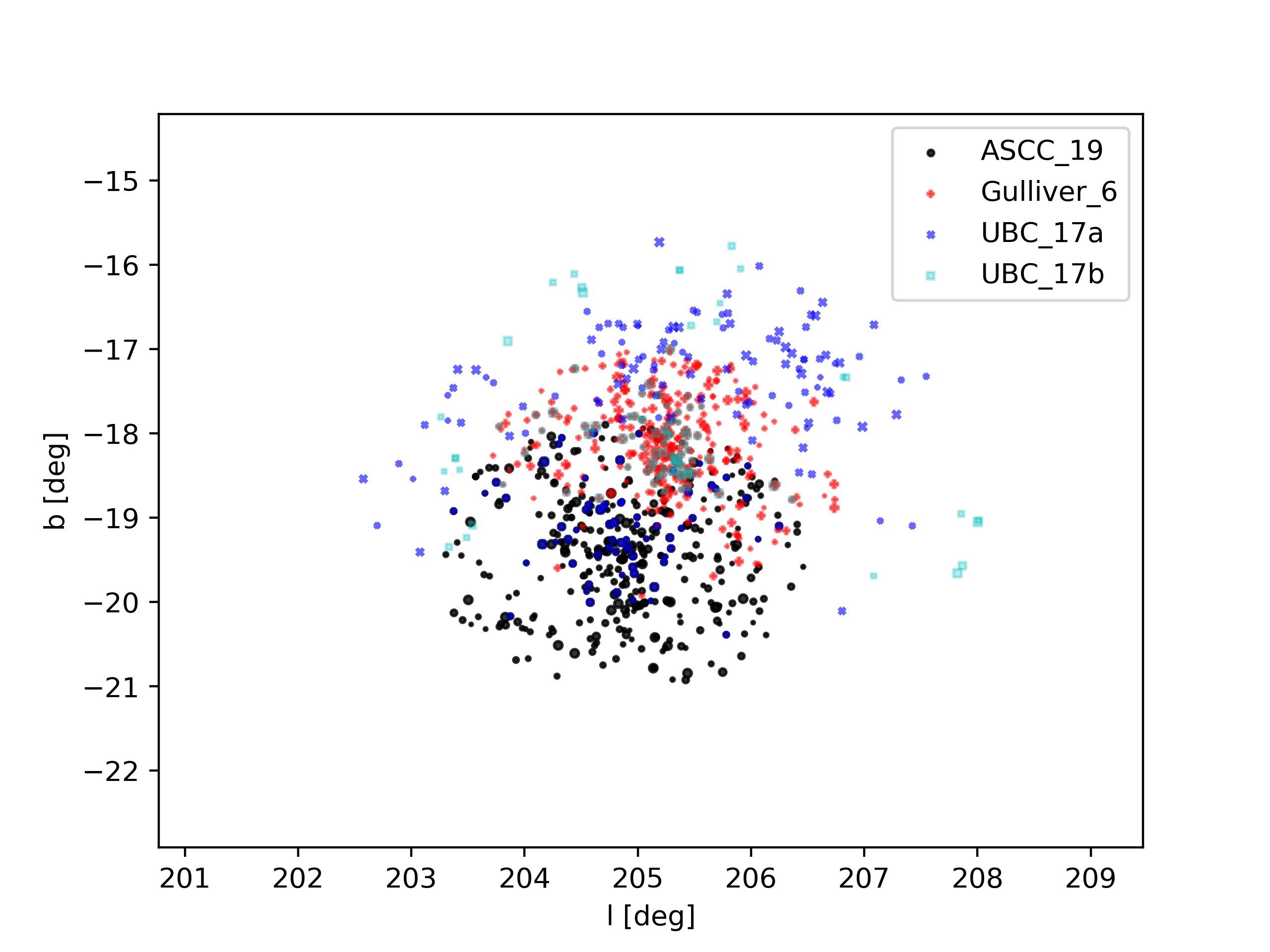}} \quad &
	 \subfloat{\includegraphics[scale=0.5]{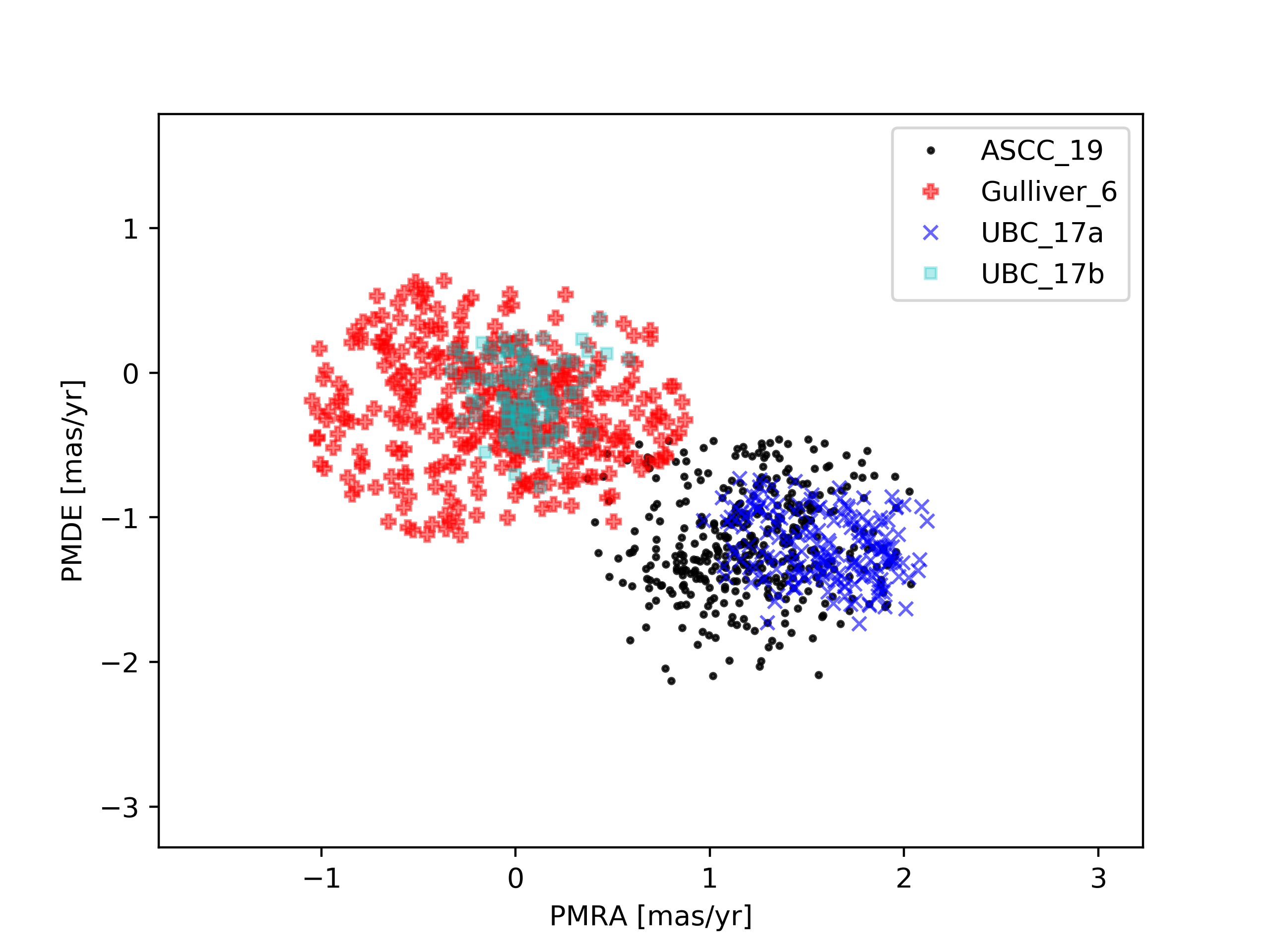}} \\
   \subfloat{\includegraphics[scale=0.5]{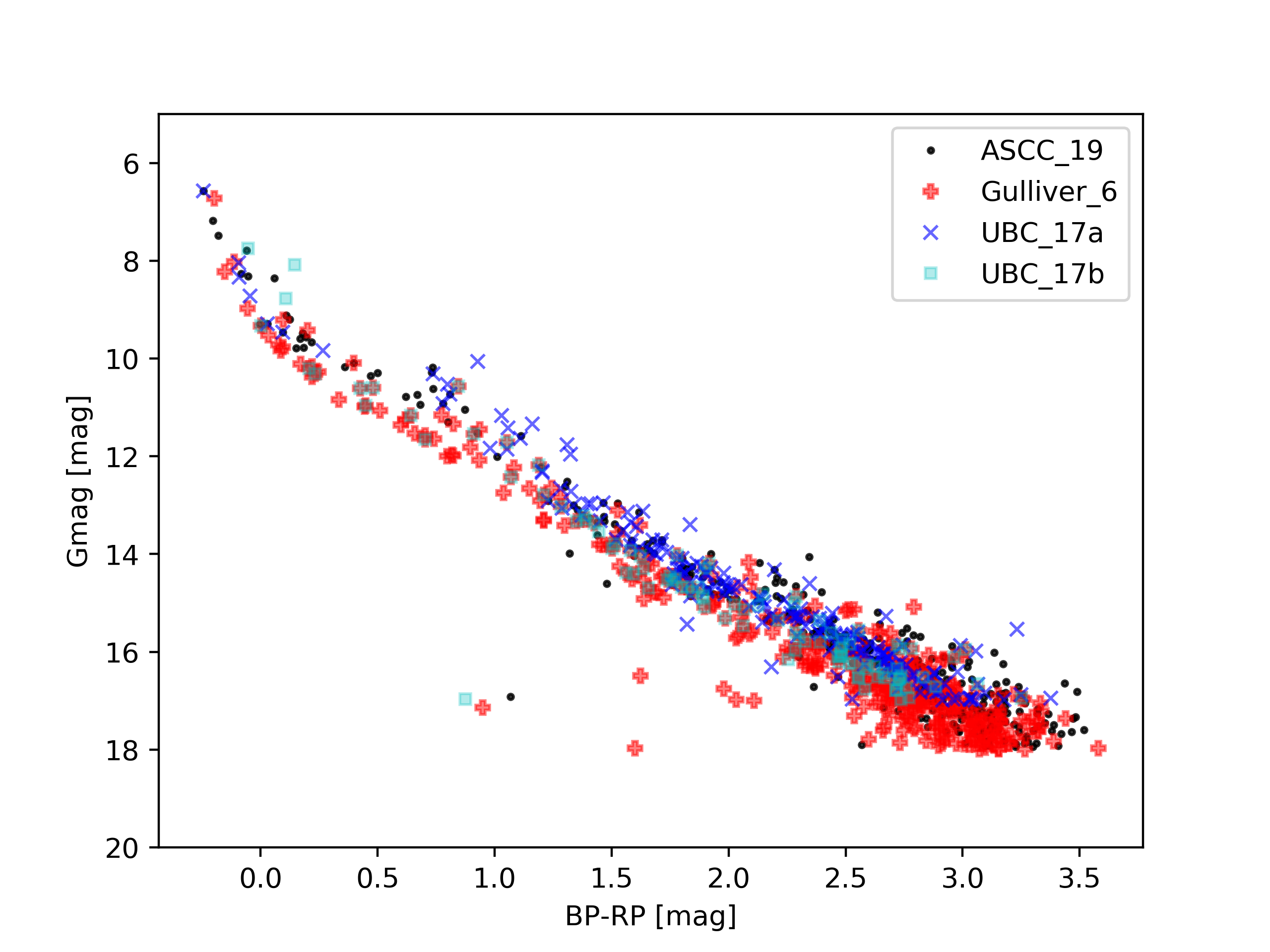}} \quad & 
   \subfloat{\includegraphics[scale=0.5]{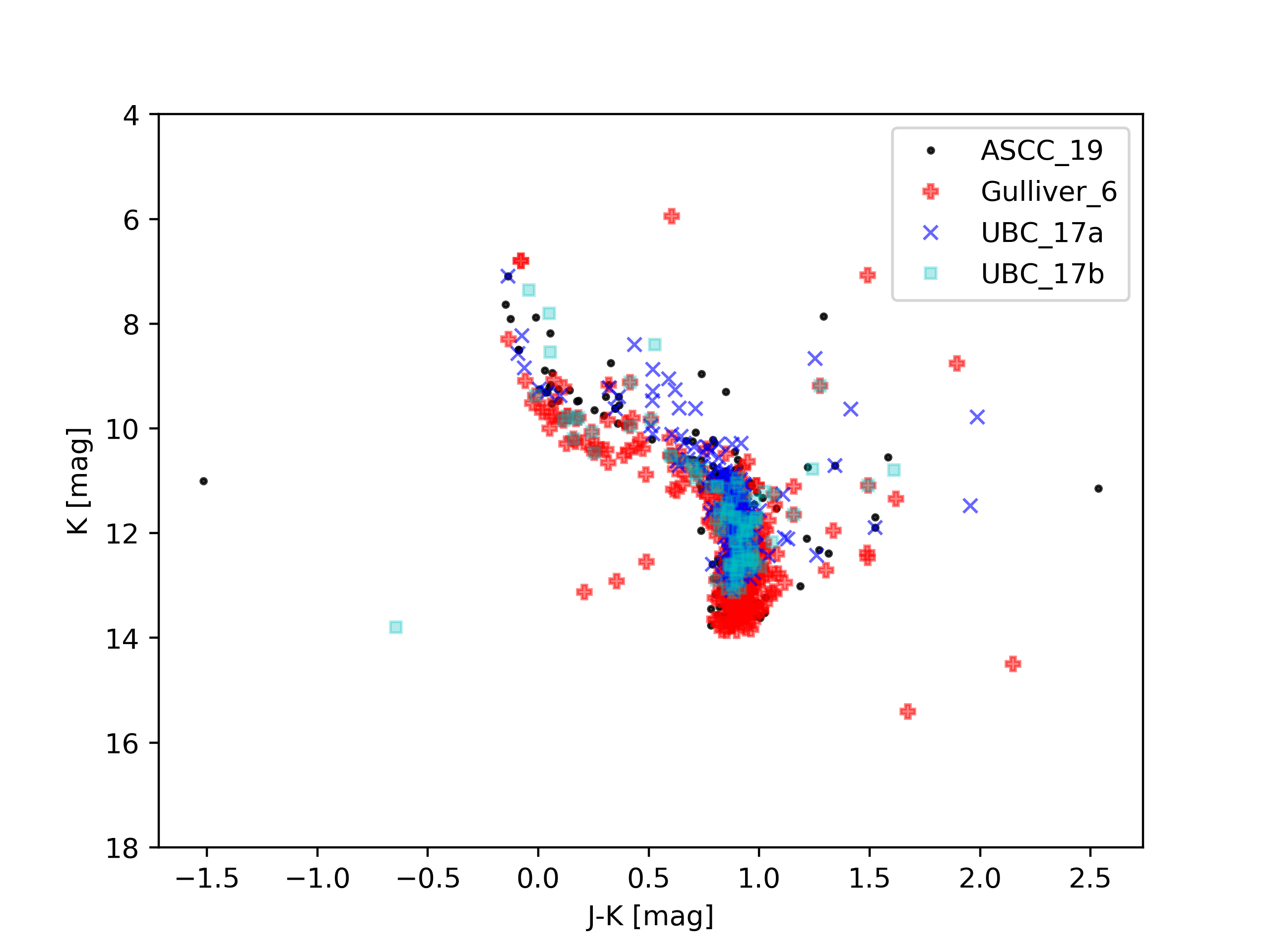}} \\
	 \end{tabular}
   \caption{Different slices through the phase space of Agg02 (ASCC~19, Gulliver~6, UBC~17a, UBC~17b), together with the complete colour-magnitude diagrams (based on Gaia and 2MASS photometry). Top-left: distribution of parallaxes of all individual stars in Agg02. Duplicates are excluded. Top-right: the histogram of parallaxes, excluding the duplicate cluster members. The best fit was achieved with the following double-Gaussian function parameters: $\varpi_1 = 2.364 \pm 0.007$ mas, $\sigma_1 = 0.134 \pm 0.009$ mas, $\varpi_2 = 2.377 \pm 0.008$ mas, $\sigma_2 = 0.018 \pm 0.011$ mas. Middle-left: coordinates of the stars in Agg02. Size of the points indicates values of the observed magnitude $G$. Middle-right: proper motion diagram of Agg02. Bottom-left: colour-magnitude diagram of Agg02, based on the observed values of $B_P-R_P$ and $G$). Bottom-right: colour-magnitude diagram of Agg02 based on 2MASS photometry. The stars were located using the coordinates in Gaia database.}
   \label{phasespace_3}
\end{figure*}

\newpage

\begin{figure*}[t]
   \centering
	 \begin{tabular}{ll}
   \multicolumn{2}{c}{\subfloat{\includegraphics[scale=0.5]{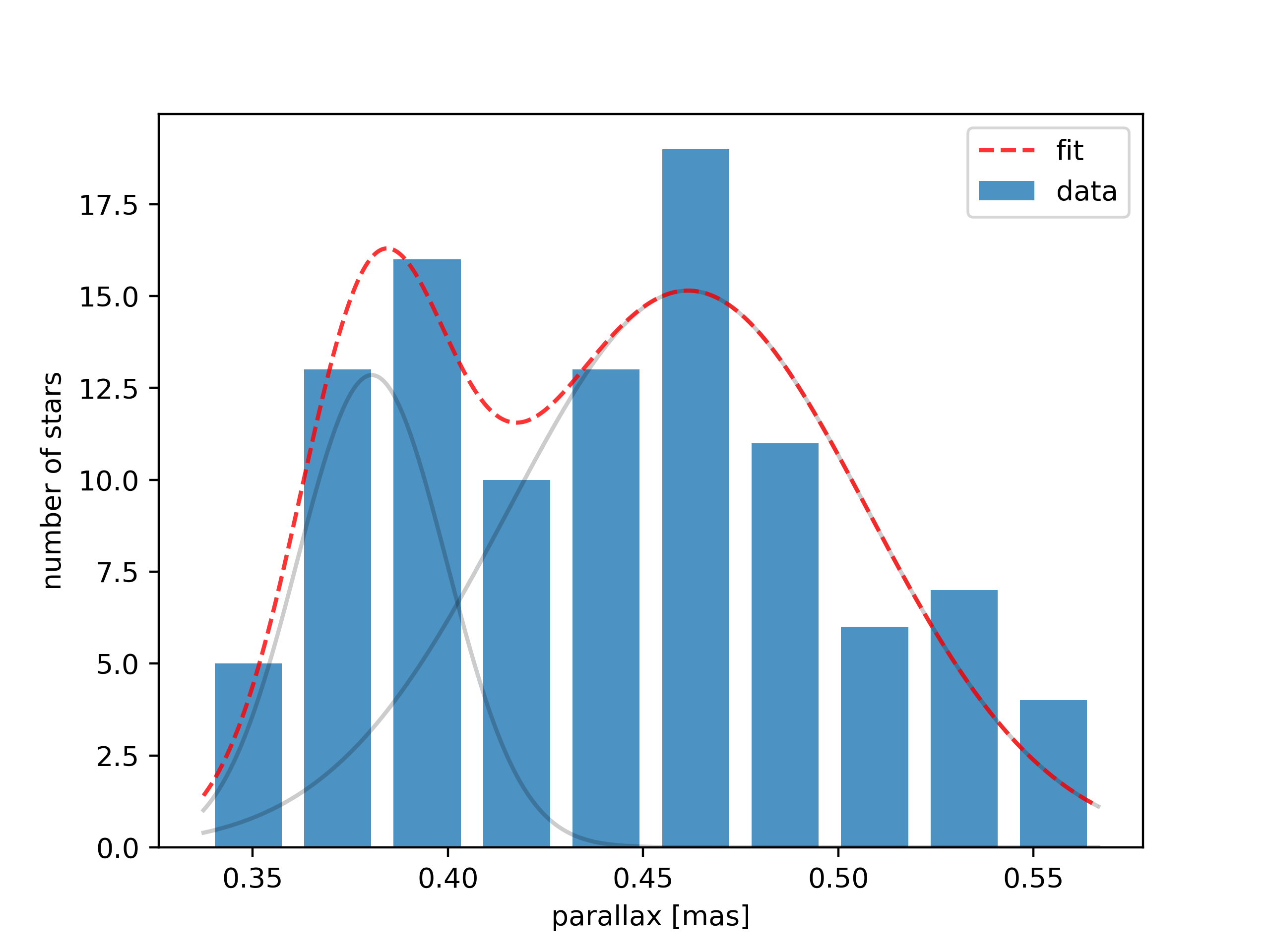}}} \\ 
   \subfloat{\includegraphics[scale=0.5]{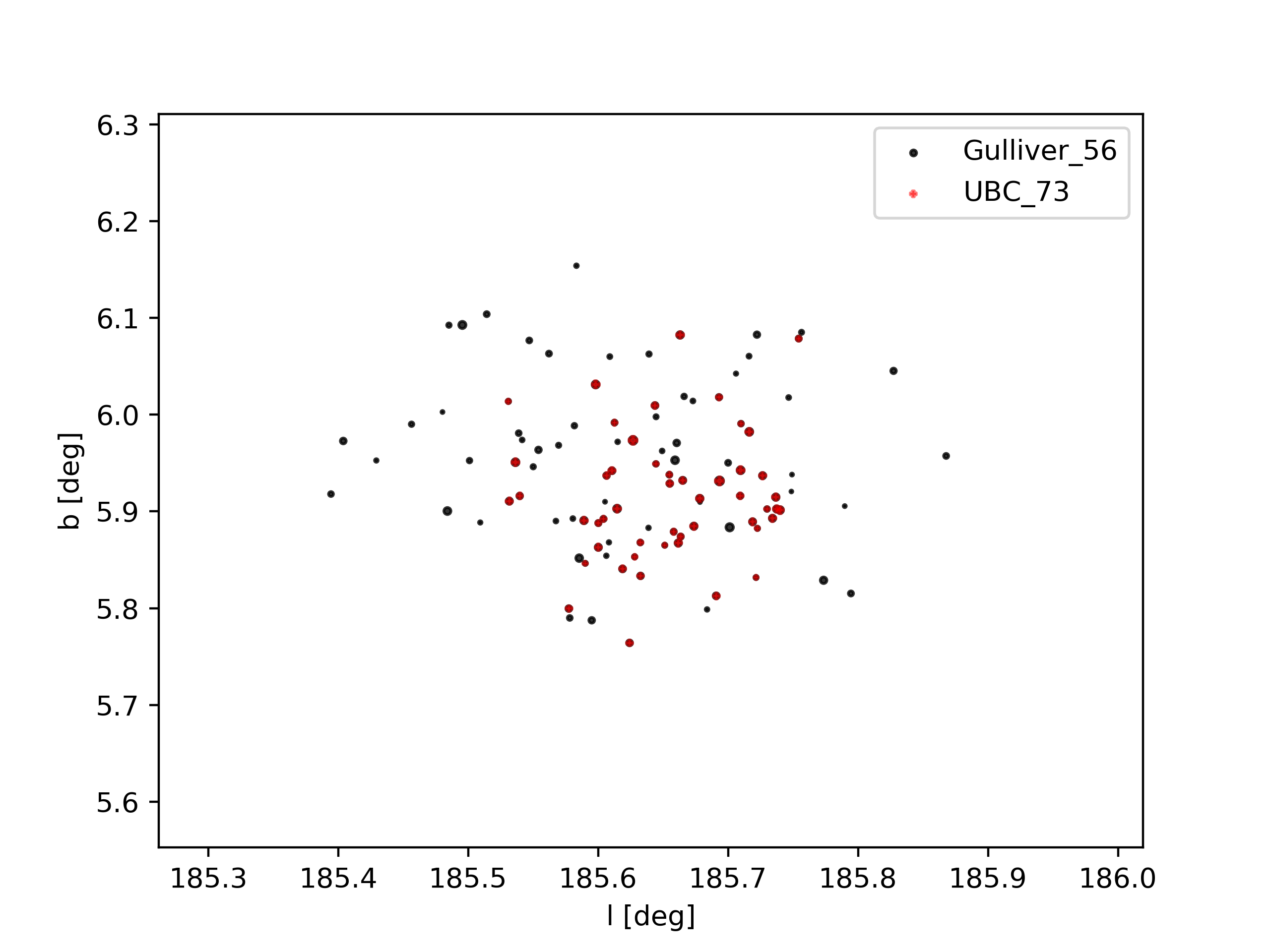}} \quad &
	 \subfloat{\includegraphics[scale=0.5]{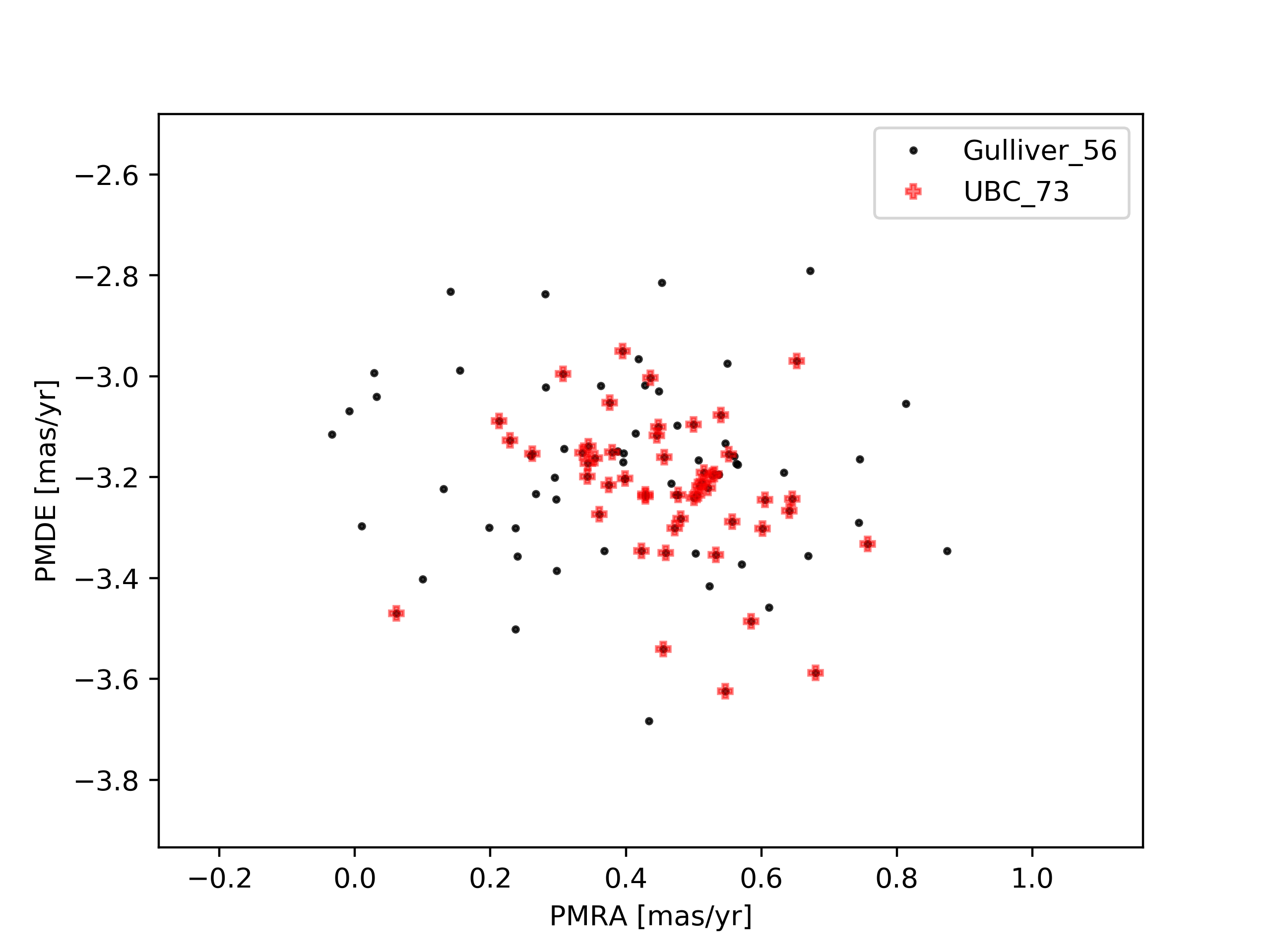}} \\
   \subfloat{\includegraphics[scale=0.5]{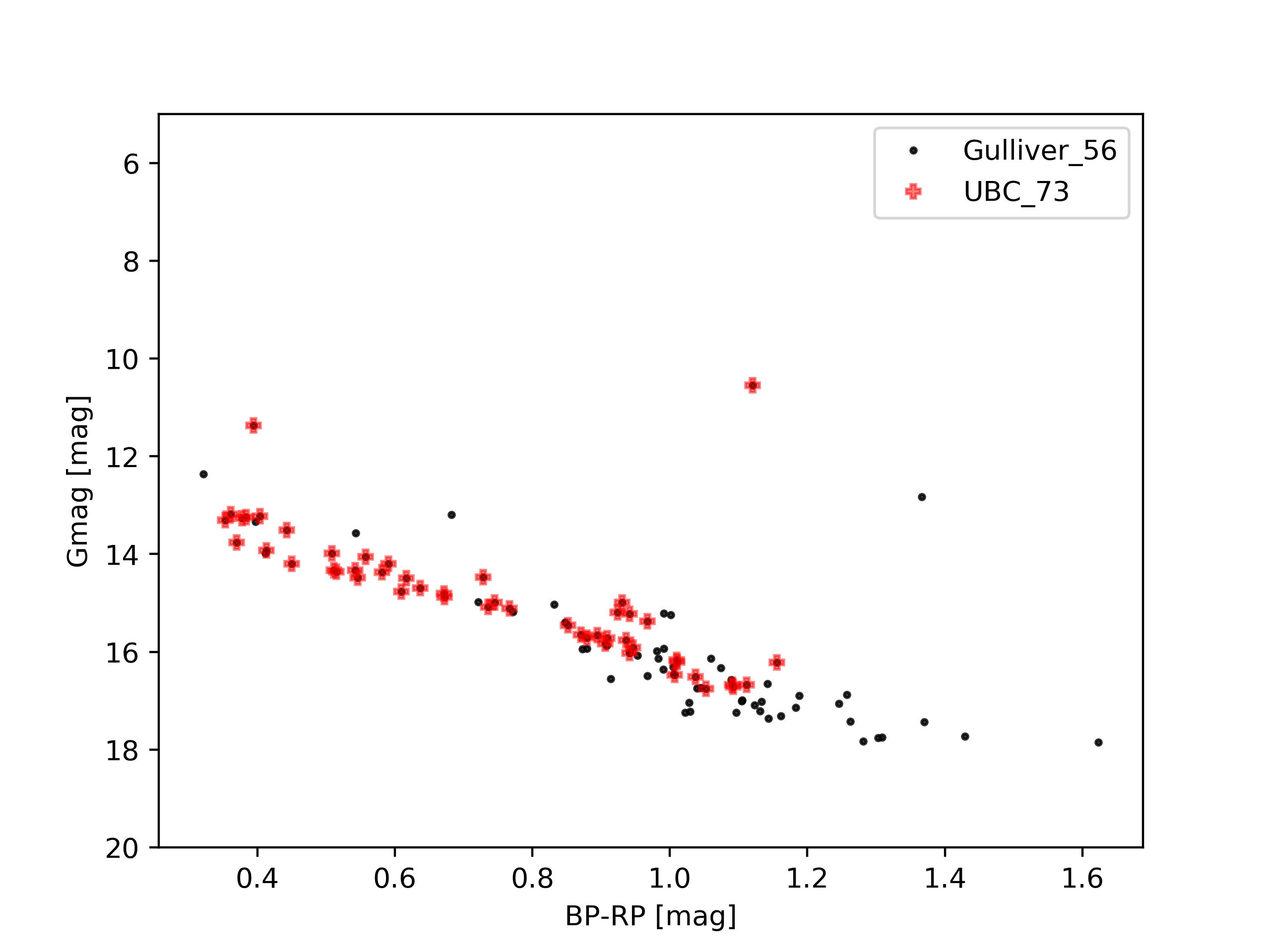}} \quad & 
   \subfloat{\includegraphics[scale=0.5]{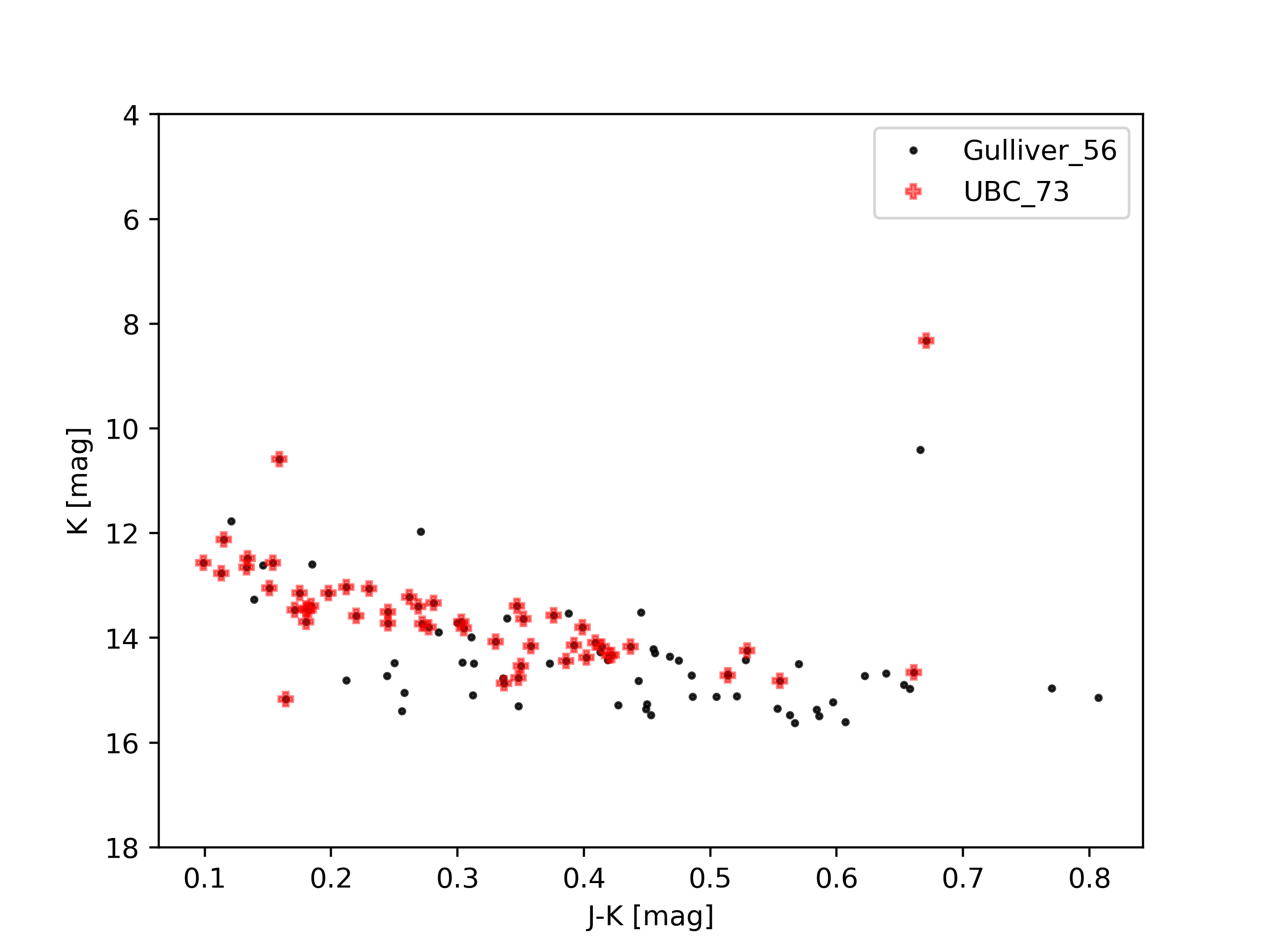}} \\
	 \end{tabular}
   \caption{Different slices through the phase space of Agg42 (Gulliver~56, UBC~73), together with the complete colour-magnitude diagrams (based on Gaia and 2MASS photometry). Top: the histogram of parallaxes, excluding the duplicate cluster members. The best fit was achieved with the following double-Gaussian function parameters: $\varpi_1 = 0.381 \pm 0.007$ mas, $\sigma_1 = 0.019 \pm 0.009$ mas, $\varpi_2 = 0.462 \pm 0.014$ mas, $\sigma_2 = 0.046 \pm 0.014$ mas. Middle-left: coordinates of the stars in Agg42. Size of the points indicates values of the observed magnitude $G$. Middle-right: proper motion diagram of Agg42. Bottom-left: colour-magnitude diagram of Agg42, based on the observed values of $B_P-R_P$ and $G$). Bottom-right: colour-magnitude diagram of Agg42 based on 2MASS photometry. The stars were located using the coordinates in Gaia database.}
   \label{phasespace_4}
\end{figure*}

\newpage

\begin{figure*}[t]
   \centering
	 \begin{tabular}{ll}
   \multicolumn{2}{c}{\subfloat{\includegraphics[scale=0.5]{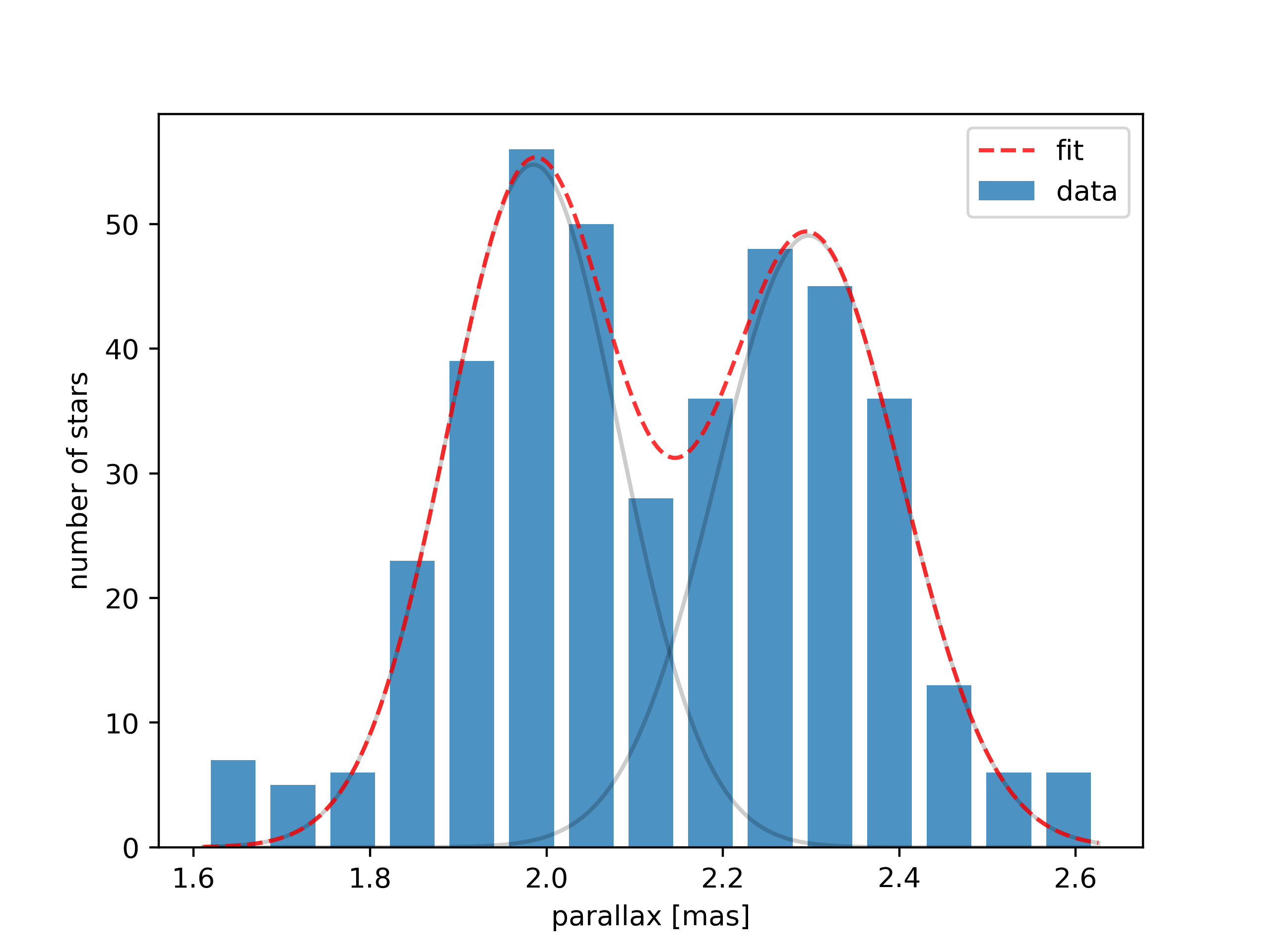}}} \\ 
   \subfloat{\includegraphics[scale=0.5]{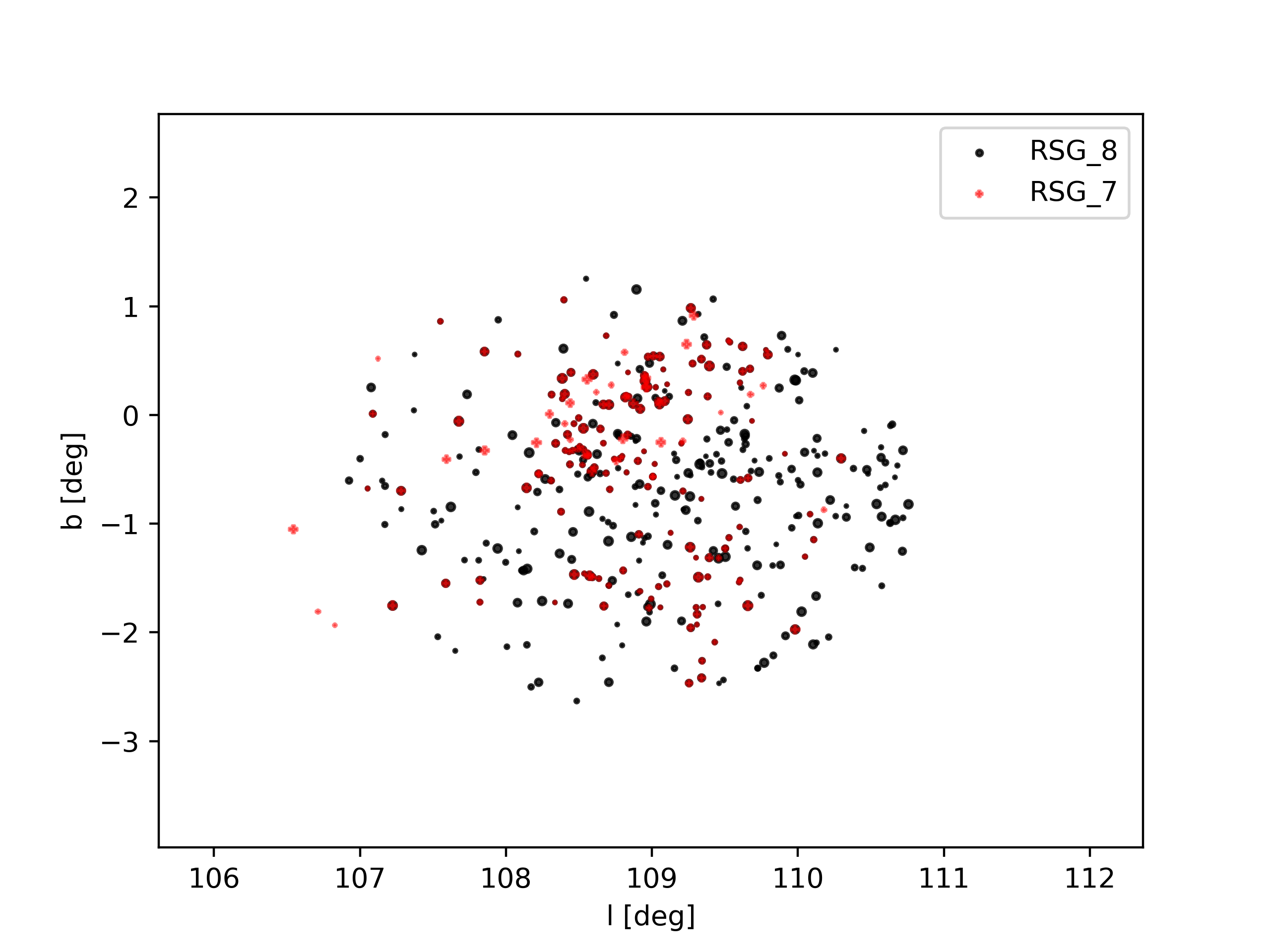}} \quad &
	 \subfloat{\includegraphics[scale=0.5]{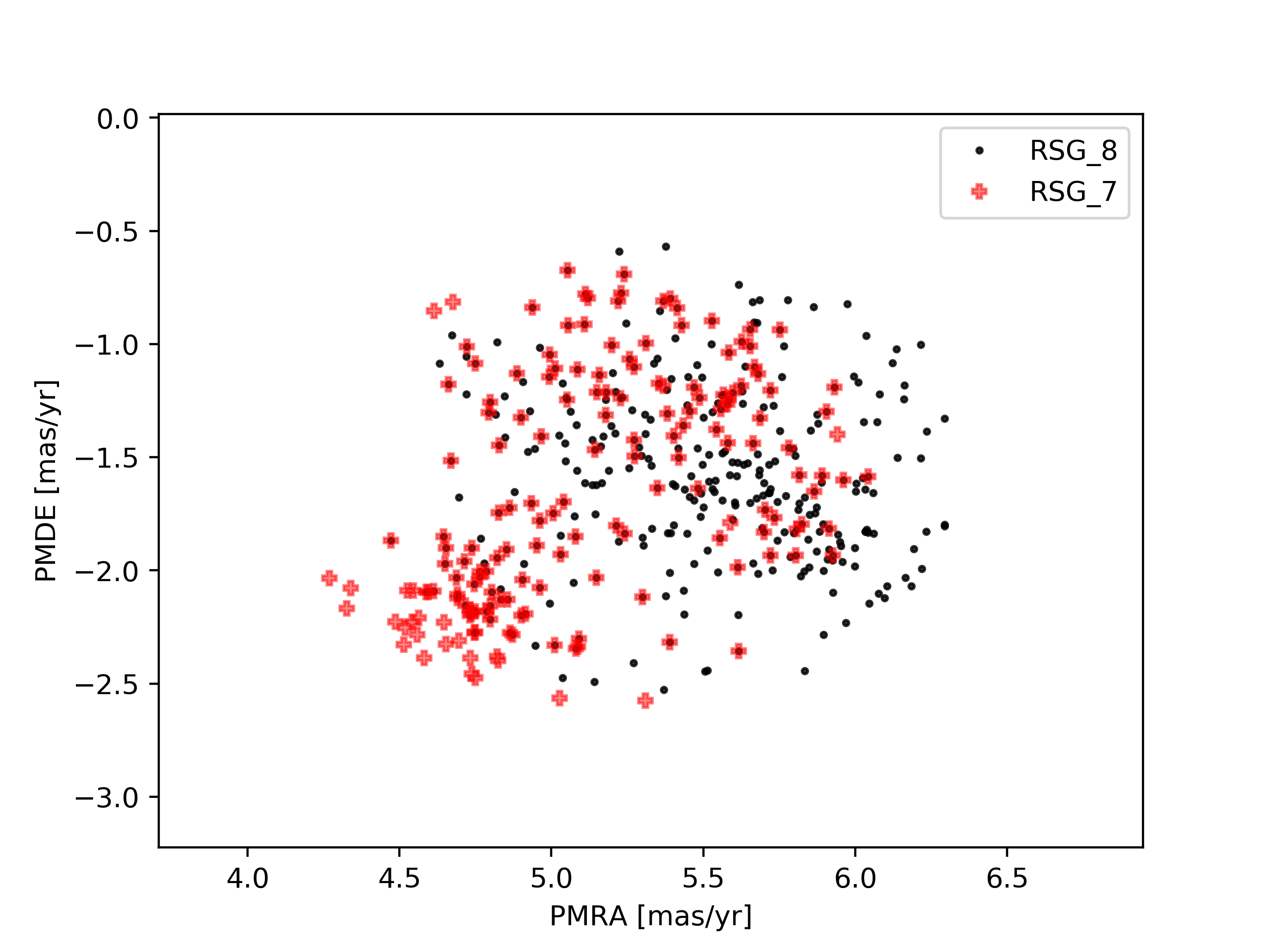}} \\
   \subfloat{\includegraphics[scale=0.5]{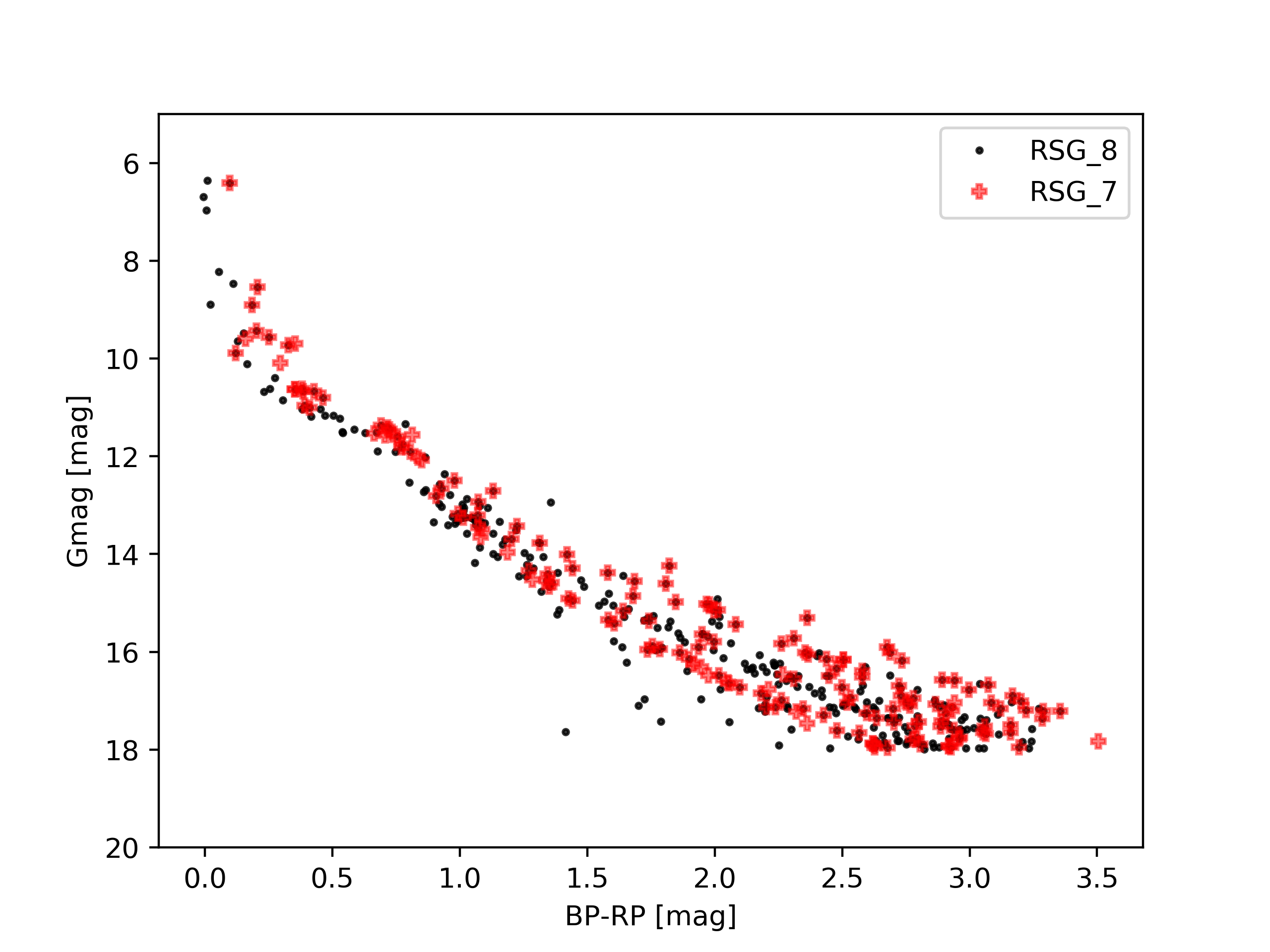}} \quad & 
   \subfloat{\includegraphics[scale=0.5]{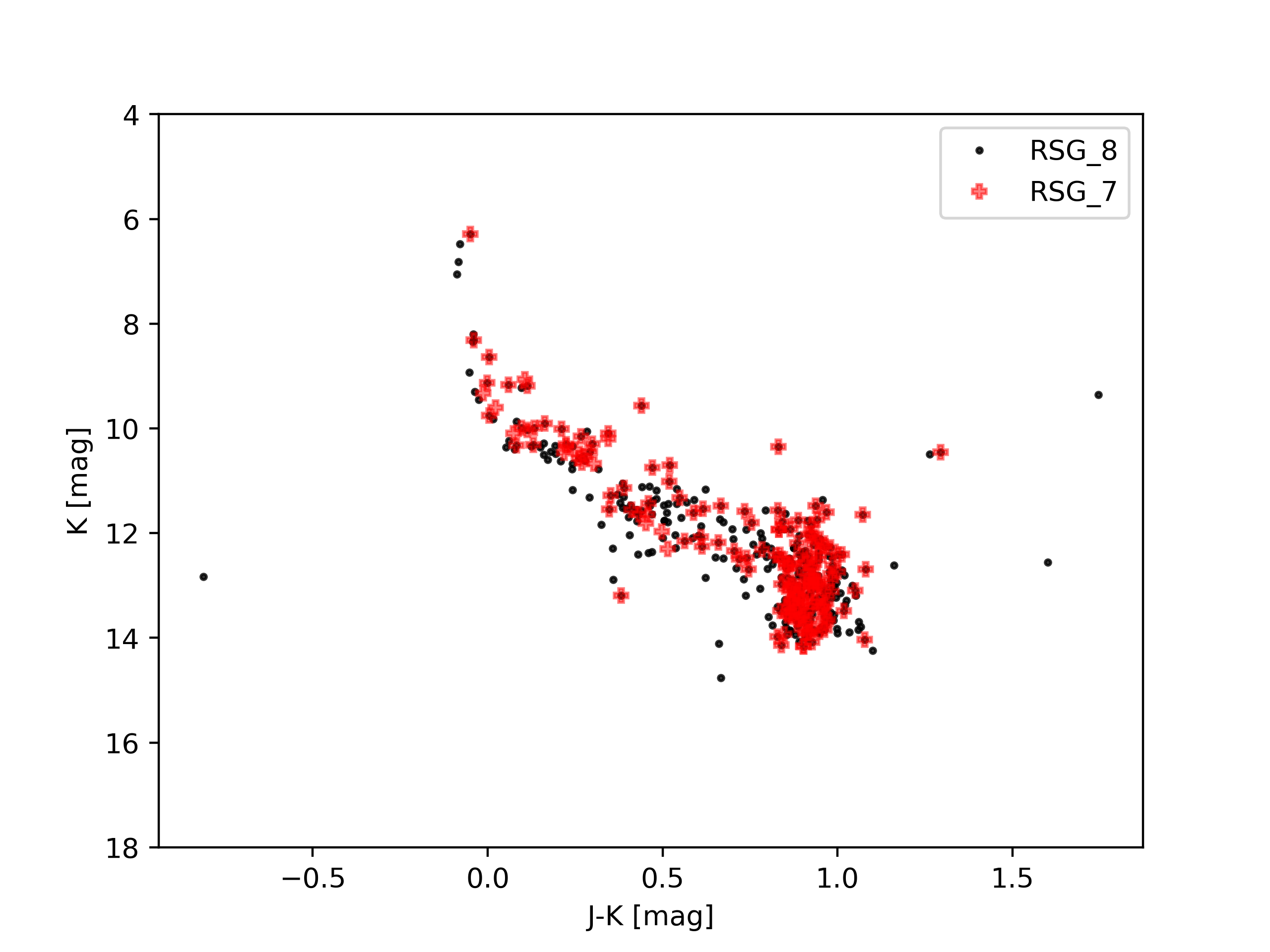}} \\
	 \end{tabular}
   \caption{Different slices through the phase space of Agg53 (RSG~7, RSG~8), together with the complete colour-magnitude diagrams (based on Gaia and 2MASS photometry). Top: the histogram of parallaxes, excluding the duplicate cluster members. The best fit was achieved with the following double-Gaussian function parameters: $\varpi_1 = 1.985 \pm 0.010$ mas, $\sigma_1 = 0.098 \pm 0.010$ mas, $\varpi_2 = 2.297 \pm 0.012$ mas, $\sigma_2 = 0.105 \pm 0.012$ mas. Middle-left: coordinates of the stars in Agg53. Size of the points indicates values of the observed magnitude $G$. Middle-right: proper motion diagram of Agg53. Bottom-left: colour-magnitude diagram of Agg53, based on the observed values of $B_P-R_P$ and $G$). Bottom-right: colour-magnitude diagram of Agg53 based on 2MASS photometry. The stars were located using the coordinates in Gaia database.}
   \label{phasespace_5}
\end{figure*}

\newpage

\begin{figure*}[t]
   \centering
	 \begin{tabular}{ll}
   \subfloat{\includegraphics[scale=0.5]{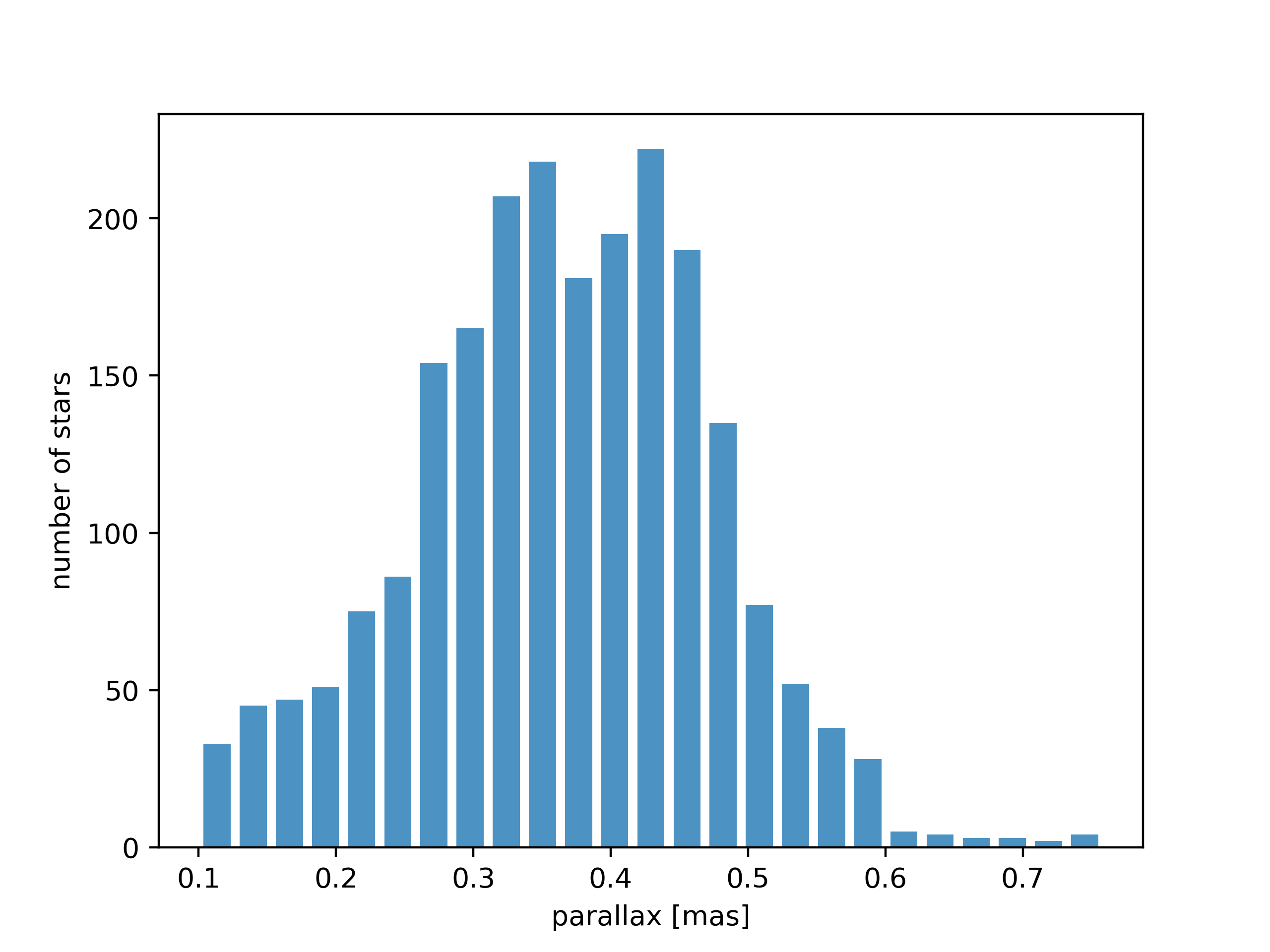}} \quad & 
   \subfloat{\includegraphics[scale=0.5]{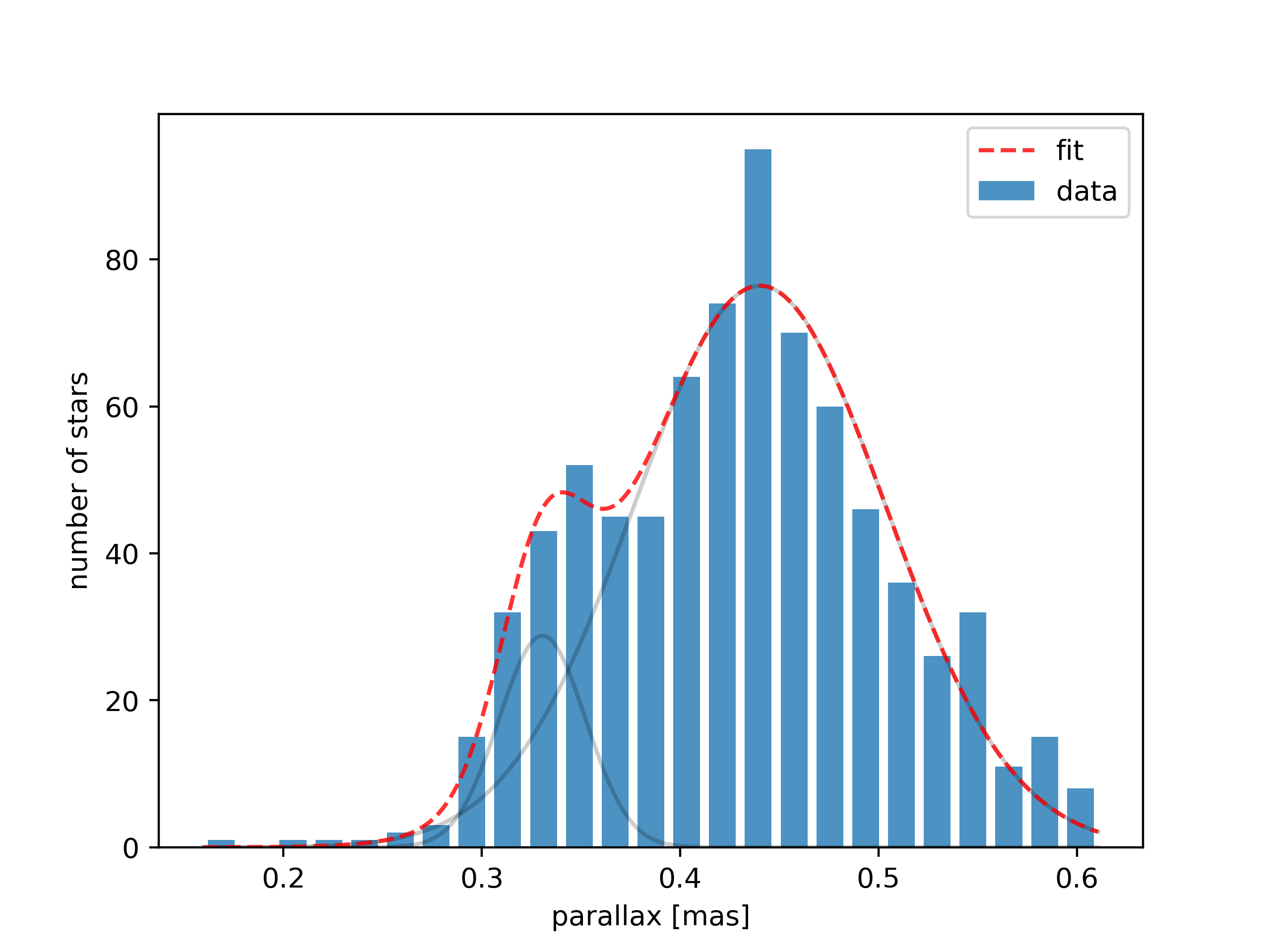}} \\
   \subfloat{\includegraphics[scale=0.5]{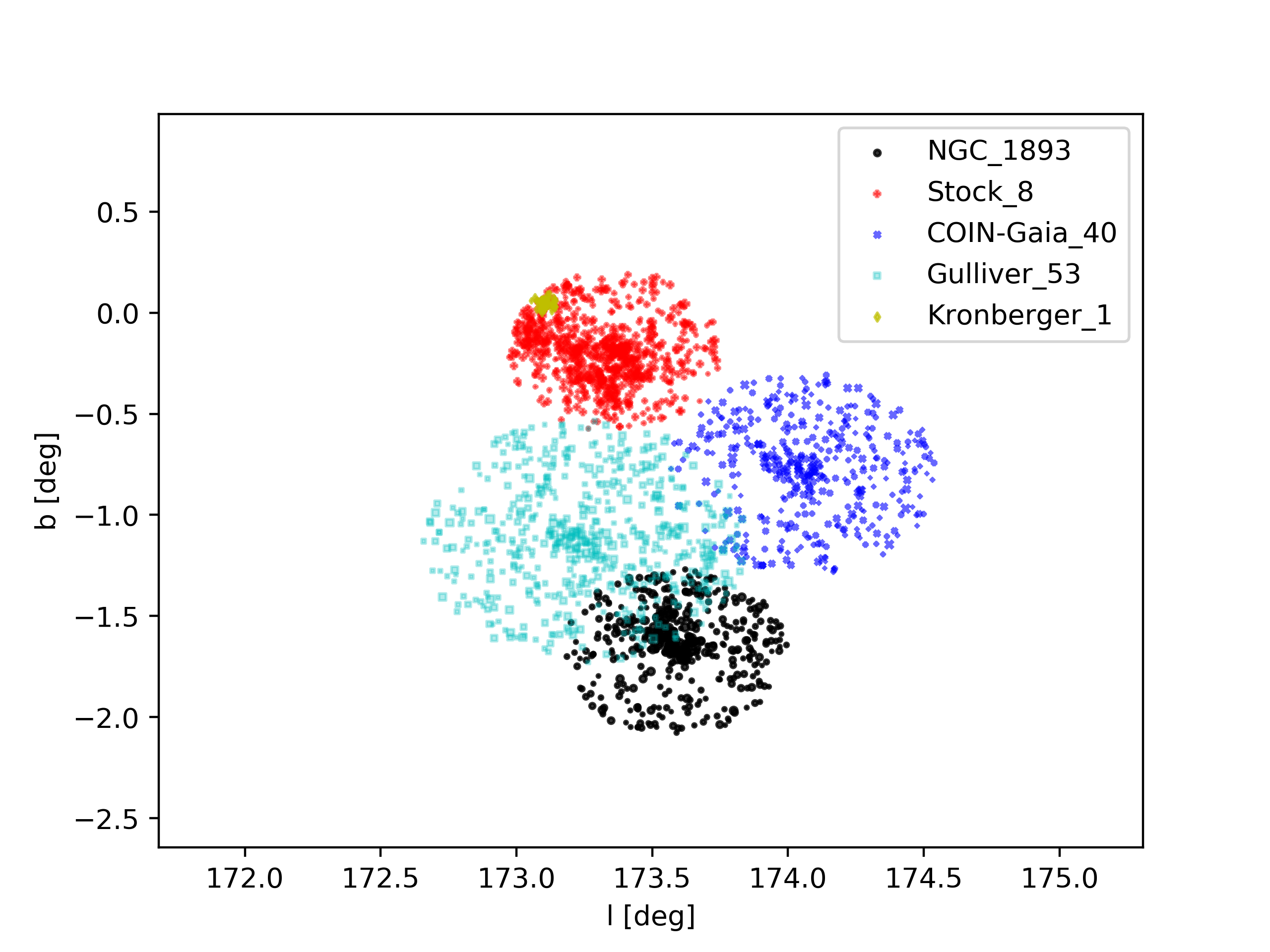}} \quad &
	 \subfloat{\includegraphics[scale=0.5]{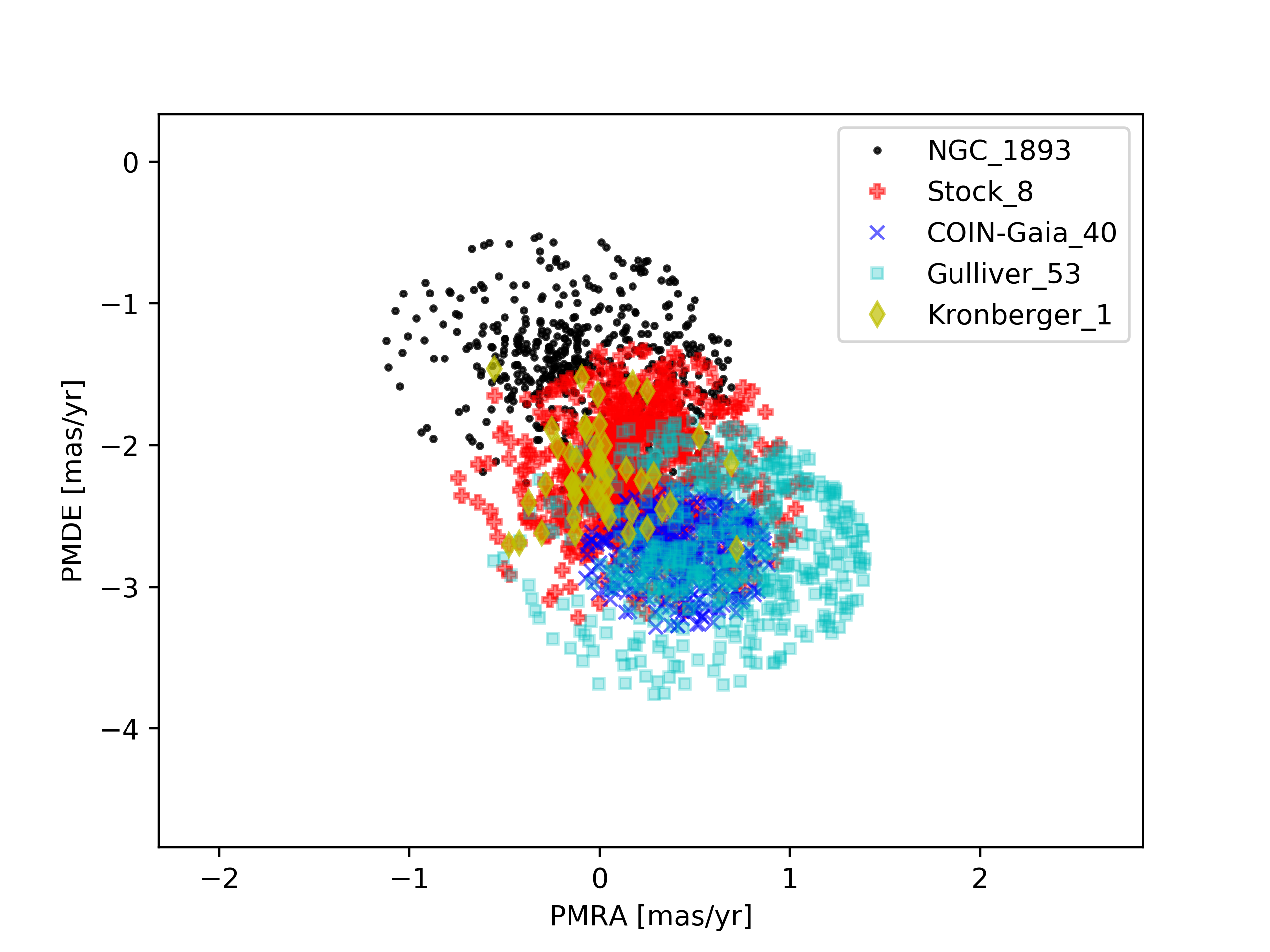}} \\
   \subfloat{\includegraphics[scale=0.5]{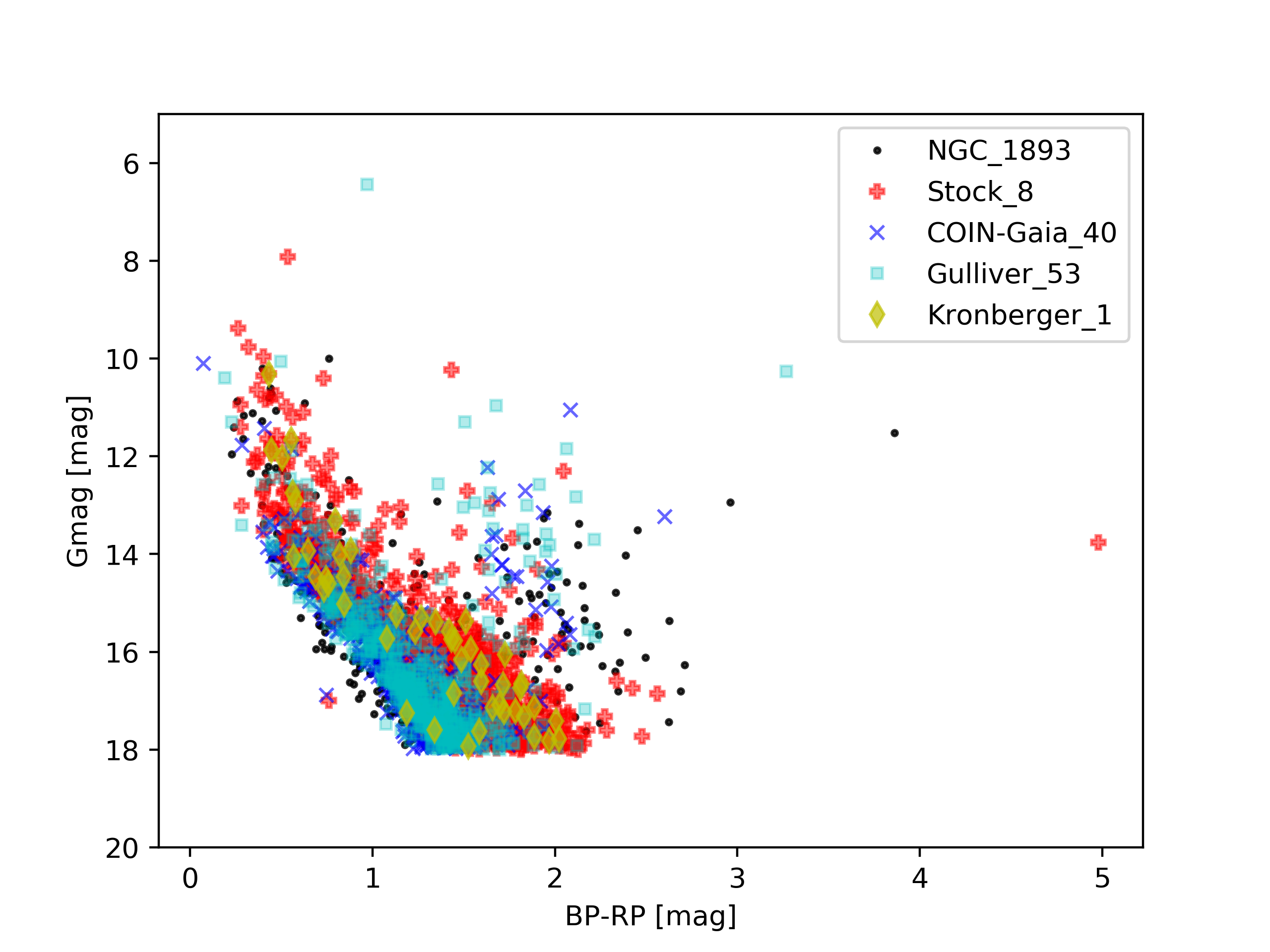}} \quad & 
   \subfloat{\includegraphics[scale=0.5]{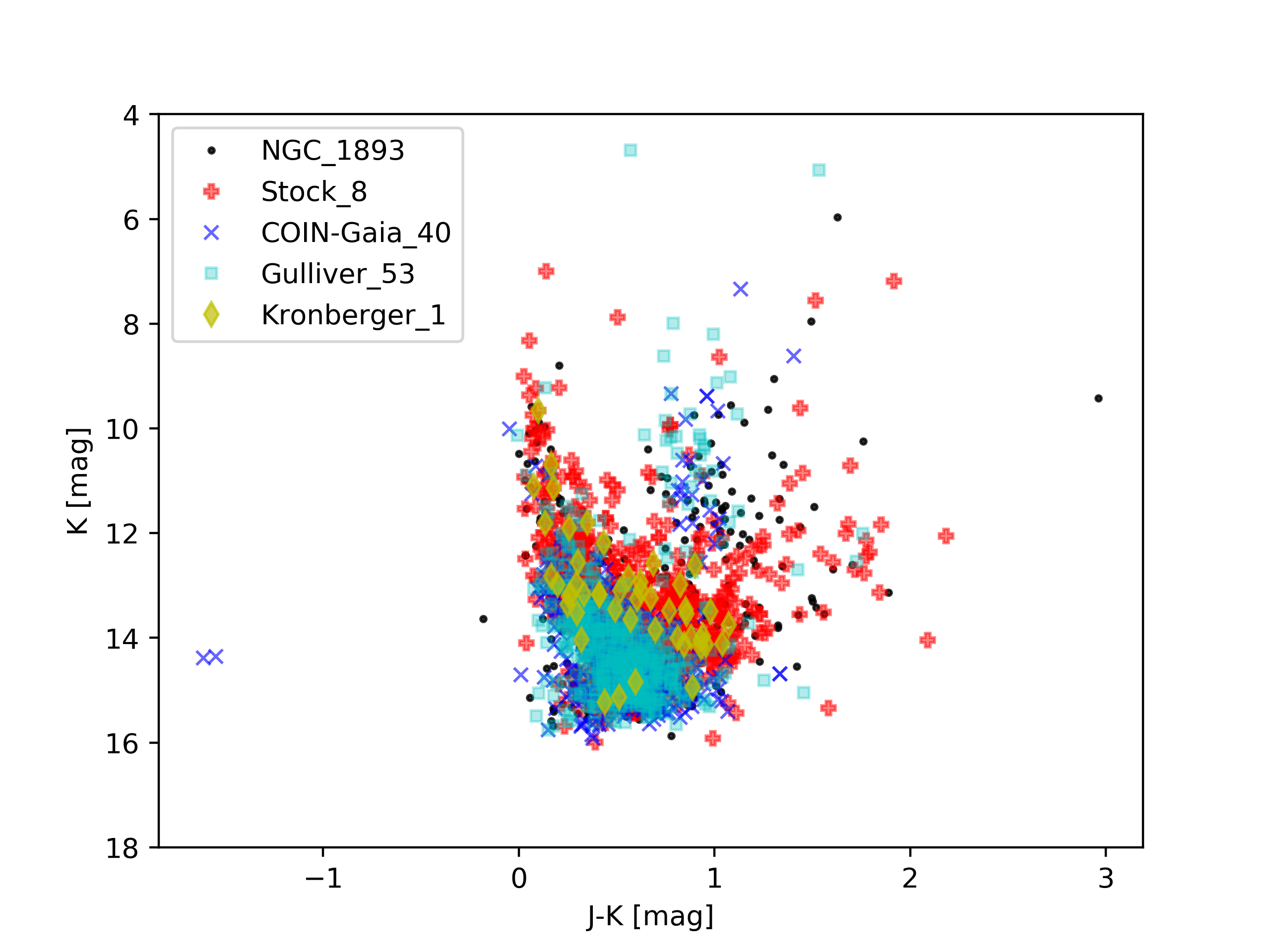}} \\
	 \end{tabular}
   \caption{Different slices through the phase space of Agg17 (COIN-Gaia~40, Gulliver~53, Kronberger~1, NGC~1893, Stock~8), together with the complete colour-magnitude diagrams (based on Gaia and 2MASS photometry). Top-left: distribution of parallaxes of all individual stars in Agg17. Duplicates are excluded. Top-right: the histogram of parallaxes, excluding the duplicate cluster members. The best fit was achieved with the following double-Gaussian function parameters: $\varpi_1 = 0.331 \pm 0.006$ mas, $\sigma_1 = 0.022 \pm 0.007$ mas, $\varpi_2 = 0.440 \pm 0.005$ mas, $\sigma_2 = 0.064 \pm 0.005$ mas. Middle-left: coordinates of the stars in Agg17. Size of the points indicates values of the observed magnitude $G$. Middle-right: proper motion diagram of Agg17. Bottom-left: colour-magnitude diagram of Agg17, based on the observed values of $B_P-R_P$ and $G$). Bottom-right: colour-magnitude diagram of Agg17 based on 2MASS photometry. The stars were located using the coordinates in Gaia database.}
   \label{phasespace_6}
\end{figure*}

\end{appendix}

\end{document}